\documentclass{article}
\usepackage{graphicx}
\usepackage{amsmath}
\usepackage{amssymb}

\begin{document}

\title{Classical dynamics on graphs}
\author{F. Barra and P. Gaspard\\{\em Center for Nonlinear Phenomena and
Complex
Systems,} \\{\em Universit\'{e} Libre de Bruxelles, Campus Plaine C.P. 231,}\\
{\em B-1050 Brussels, Belgium.}}

\maketitle

\begin{abstract}
We consider the classical evolution of a particle on a graph by using
a time-continuous Frobenius-Perron operator which generalizes previous
propositions. In this way, the relaxation rates as well as the chaotic
properties can be defined
for the time-continuous classical dynamics on graphs.  These properties are
given as the zeros of
some periodic-orbit zeta functions.  We consider in detail the case of
infinite periodic graphs where
the particle undergoes a diffusion process. The infinite spatial extension
is taken into account by
Fourier transforms which decompose the observables and probability
densities into sectors
corresponding to different values of the wave number. The hydrodynamic
modes of diffusion are
studied by an eigenvalue problem of a Frobenius-Perron operator
corresponding to a given sector. The
diffusion coefficient is obtained from the hydrodynamic modes of diffusion
and has the
Green-Kubo form. Moreover, we study finite but large open graphs
which converge to the infinite periodic graph when their size goes to
infinity. The lifetime of the particle on the open graph
is shown to correspond to the lifetime of a system which undergoes a
diffusion process
before it escapes.

03.65.Sq (Semiclassical theories and applications);

05.60.Cd	(Classical transport);

45.05.+x	(General theory of classical mechanics of discrete systems).
\end{abstract}

\section{Introduction}

The study of classical dynamics on graphs is motivated by the recent
discovery that quantum graphs have similar spectral statistics of
energy levels as the classically chaotic quantum systems
\cite{Smilansky0,Smilansky1}.
Since this pioneering work by Kottos and Smilansky, several studies have
been devoted to the spectral properties of quantum graphs
\cite{Smilansky2,Keating,Felipe3}.
However, the classical dynamics, which is of great importance for the
understanding of the short-wavelength quantum properties, has not yet been
considered in detail.
In Refs. \cite{Smilansky0,Smilansky1}, a classical dynamics has been
considered in
which the particles are supposed to move on the graph with a discrete and
isochronous
(topological) time, ignoring the different lengths of the bonds composing
the graph.

The purpose of the present paper is to develop the theory of the classical
dynamics on
graphs by considering the motion of particles in real time.  This
generalization is
important if we want to compare the classical and quantum quantities,
especially, with regard to
the time-dependent properties in open or spatially extended graphs.  A
real-time classical
dynamics on graphs should allow us to define kinetic and transport
properties such as the
classical escape rates and the diffusion coefficients, as well as the
characteristic quantities
of chaos such as the Kolmogorov-Sinai (KS) and the topological entropies
per unit time.

An important question concerns the nature of the classical dynamics on a graph.
A graph is a network of bonds on which the classical particle has a
one-dimensional
uniform motion at constant energy.  The bonds are interconnected at
vertices where several bonds
meet.  The number of bonds connected with a vertex is called the valence of
the vertex.
A quantum mechanics has been defined on such graphs by considering a
wavefunction
extending on all the bonds \cite{Smilansky0,Smilansky1}. This wavefunction
has been supposed to
obey the one-dimensional Schr\"odinger equation on each bond.  The
Schr\"odinger equation is
supplemented by boundary conditions at the vertices.  The boundary
conditions at a vertex determine
the quantum amplitudes of the outgoing waves in terms of the amplitudes of
the ingoing waves and,
thus, the transmission and reflection amplitudes of that particular vertex.

In the classical limit, the Schr\"odinger equation leads to Hamilton's
classical equations
for the one-dimensional motion of a particle on each bond.  When a vertex
is reached, the square
moduli of the quantum amplitudes give the probabilities that the particle
be reflected
back to the ingoing bond or be transmitted to one of the other bonds
connected with the
vertex.  In the classical limit of arbitrarily short wavelengths, the
transmission and reflection
probabilities do not reduce to the trivial ones (i.e., to zero and one) for
typical graphs.
Accordingly, the limiting classical dynamics on graphs is in general a
combination of the uniform
motion of the particle on the bonds with random transitions at the
vertices.  This dynamical
randomness which naturally appears in the classical limit is at the origin
of a splitting of
the classical trajectory into a tree of trajectories.  This feature is not
new and has already
been observed in several processes such as the ray splitting in billiards
divided by a potential
step \cite{bluemel} or the scattering on a wedge \cite{griffiths}.  We
should emphasize that this
dynamical randomness manifests itself only on subsets with a dimension
lower than the phase space
dimension and not in the bulk of phase space, so that the classical graphs
share many properties of
the deterministic chaotic systems, as we shall see below.

The dynamical randomness of the classical dynamics on graphs requires a
Liouvillian
approach to describe the time evolution of the probability density to find
the particle somewhere
on the graph.  Accordingly, one of our first goals below will be to derive
the Frobenius-Perron
operator as well as the associated master equation for the graphs.  This
operator is introduced
by noticing that the classical dynamics on a graph is equivalent to a
random suspended flow
determined by the lengths of the bonds, the velocity of the particle, and
the transition
probabilities.

A consequence of the dynamical randomness is the relaxation of the
probability density toward the
equilibrium density in typical closed graphs, or to zero in open graphs or
in graphs of infinite
extension.  This relaxation can be characterized by the decay rates which
are given by solving
the eigenvalue problem of the Frobenius-Perron operator.  The
characteristic determinant of the
Frobenius-Perron operator defines a classical zeta function and its zeros
-- also called the
Pollicott-Ruelle resonances -- give the decay rates.  The leading decay
rate is the so-called
escape rate.  The Pollicott-Ruelle resonances have a particularly important
role to play because
they control the decay or relaxation and they also manifest themselves in
the quantum scattering
properties of open systems, as revealed by a recent experiment by Sridhar
and coworkers
\cite{Sridhar}.   The decay rates are time-dependent properties so that
they require to consider the
continuous-time classical dynamics to be defined.

Besides, we define a continuous-time ``topological pressure" function from
which the different
chaotic properties of the classical dynamics on graphs can be deduced.
This function allows us
to define the KS and topological entropies per unit time, as well as an
effective positive Lyapunov
exponent for the graph.

We shall also show how diffusion can be studied on spatially periodic
graphs thanks to our
Frobenius-Perron operator and its decay rates.  Here, we consider graphs
that are constructed
by the repetition of a unit cell. When the cell is repeated an infinite
number of times we form a periodic graph. Such spatially extended periodic
systems are
interesting for the study of transport properties. In fact at the classical
level it has been shown in several works that relationships exist between
the chaotic dynamics
and the normal transport properties such as diffusion \cite{GaspEigenDiff}
and the thermal
conductivity \cite{alonso-casati}, which have been studied in the periodic
Lorentz gas.
In the present paper, we obtain the continuous-time diffusion properties
for the spatially
periodic graphs.  Moreover, we also study the escape rate in large but
finite open graphs
and we show that this rate is related, on the one hand, to the diffusion
coefficient and, on the
other hand, to the effective Lyapunov exponent and the KS entropy per unit
time.

The plan of the paper is the following.
Sec. \ref{sec.graphs} contains a general introduction to the graphs and
their classical dynamics.
In Subsec. \ref{subsec.susp}, we introduce the evolution using the
aforementioned random
suspended flow and, therefore, we can follow the approach developed in Ref.
\cite{GaspEigenDiff}
for the study of relaxation and chaotic properties at the level of the
Liouvillian dynamics which
is developed in Sec. \ref{sec.liouville}.  The Frobenius-Perron operator is
derived in Subsec.
\ref{subsec.FP}.  In Subsec. \ref{subsec.master}, we present an alternative
derivation of the
Frobenius-Perron operator and its eigenvalues and eigenstates, which is
based on a
master-equation approach, familiar in the context of stochastic processes.
Both approaches are
shown to be equivalent.  In Sec. \ref{sec.relax}, we study the relaxation
and ergodic properties
of the graphs in terms of the classical zeta function and its
Pollicott-Ruelle resonances. In
Sec. \ref{sec.chaotic}, the large-deviation formalism is introduced which
allows us to
characterize the chaotic properties of these systems. In Sec.
\ref{sec.scatt}, the theory is applied to classical scattering on open
graphs.  The case of
infinite periodic graphs is considered in Sec. \ref{sec.diff}, where we
obtain the diffusion
coefficient and we show that it can be written in the form of a Green-Kubo
formula. In Sec.
\ref{sec.esc.diff}, we consider finite open graphs of the scattering type,
where the particle escape
to infinity, and we show how the diffusion coefficient can be related to
the escape rate and the
chaotic properties.  The case of infinite disordered graphs is considered
in Sec.
\ref{sec.disorder}.  Conclusions are drawn in Sec.
\ref{conclu}.

\section{The graphs and their classical dynamics}

\label{sec.graphs}

\subsection{Definition of the graphs}

\label{subsec.geom}

Now, let us introduce graphs as geometrical objects where a particle moves.
Graphs are $V$ vertices connected by $B$ bonds.
Each bond $b$ connects two vertices,
$i$ and $j$. We can assign an orientation to each bond and define ``oriented
or directed bonds''. Here one fixes the direction of the bond $[i,j]$ and
call $b=(i,j)$ the bond oriented from $i$ to $j$. The same bond but oriented
from $j$ to $i$ is denoted $\hat{b}=(j,i)$.  We notice that
$\hat{\hat{b}}=b$.  A graph with $B$ bonds has $2B$ directed bonds.
The valence $\nu_i$ of a vertex is the number of bonds that meet at the
vertex $i$.

Metric information is introduced by assigning a length $l_b$ to each bond $b$.
In order to define the position of a particle on the graph, we introduce a
coordinate $x_b$ on each bond $b=[i,j]$. We can assign either the
coordinate $x_{(i,j)}$ or
$x_{(j,i)}$. The first one is defined such that $x_{(i,j)}=0$ at $i$ and
$x_{(i,j)}=l_b$ at
$j$, whereas $x_{(j,i)}=0$ at $j$ and $x_{(j,i)}=l_b$ at $i$. Once the
orientation is
given, the position of a particle on the graph is determined by the
coordinate $x_{b}$ where $0\leq x_{b}\leq l_{b}$. The index $b$ identifies
the bond and the value of $x_{b}$ the position on this bond.

For some purposes, it is convenient to consider $b$ and $\hat{b}$ as
different bonds within the formalism. Of course, the physical quantities
defined on each of them must satisfy some consistency relations. In particular,
we should have that $l_{\hat{b}}=l_{b}$ and $x_{\hat{b}}=l_{b}-x_{b}$.

A particle on a graph moves freely as long as it is on a bond. The vertices
are singular points, and it is not possible to write down the analogue of
Newton's equations at the vertices. Instead we have to introduce transition
probabilities from bond to bond. These transition probabilities
introduce a dynamical randomness which is coming from the quantum dynamics
in the
classical limit. In this sense, the classical dynamics on graphs turns out
to be
intrinsically random.

The reflection and transmission (transition) probabilities are determined
by the quantum dynamics
on the graph. This latter introduces the probability amplitudes
$T_{bb'}$ for a transition from the bond $b'$ to the bond $b$.
We shall show in a separate paper \cite{Felipe4} that the random
classical dynamics defined in the present paper, with
the transition probabilities defined by $P_{bb'}=|T_{bb'}|^2$
is, indeed, the classical limit of the quantum dynamics on graphs.
For example, we may consider a quantum graph with transition amplitudes of
the form
\begin{equation}
T_{bb'}=C_{bb'}\left(\frac{2}{\nu_{bb'}}-\delta_{\hat{b'}b'}\right)
\label{Ts}
\end{equation}
where $C_{bb'}$ is 1 if the bond $b'$ is connected with the bond $b$
and zero otherwise and $\nu_{bb'}$ is the valence of the vertex that connects
$b'$ with $b$.
Such probability amplitudes are obtained once we impose
the continuity of the wave function and
the current conservation at each vertex.

For the classical dynamics on graphs, the energy of the particle is
conserved during the free motion in the bonds and also in the transition
to other bonds. Accordingly, we consider
the surface of constant energy in the phase space determined by the
coordinate of the particle, that is $x_{b}$ which specifies a bond and the
position with respect to a vertex. The momentum is simply given by the
direction in which the particle moves on the bond. The modulus of the
momentum is fixed by the energy. We see that position and
direction can be combined together if we speak about position in a given
directed bond. In this way our phase space are the points of the directed
bonds.
The equation of motion is thus
\begin{equation}
\frac{dx}{dt} = v = \sqrt{2E/M} \ , \qquad \hbox{for} \qquad 0<x=x_b<l_b \ ,
\end{equation}
where $v$ is the velocity in absolute value, $E$ the energy, and $M$ the
mass of the particle.
When the particle reaches the end $x_{b^{\prime }}=l_{b^{\prime
}}$ of the bond $b'$ a transition can bring it at the beginning $x_{b}=0$
of the bond $b$.
According to the above discussion, we assume moreover that this transition
from the
bond $b^{\prime}$ to the bond $b$ has the probability $P_{bb^{\prime }}$ to
occur:
\begin{equation}
\mbox{transition} \quad b^{\prime}\to b \quad \mbox{with probability}\quad
P_{bb^{\prime }}
\end{equation}
By the conservation of the total probability, the transition probabilities
must satisfy
\begin{equation}
\sum_b P_{bb^{\prime }} = 1 \label{norm}
\end{equation}
which means that the vector $\lbrace 1,1,\ldots,1 \rbrace$ is always a left
eigenvector with
eigenvalue
$1$ for the transition matrix $\mathsf P=\lbrace P_{bb^{\prime }}
\rbrace_{b,b'=1}^{2B}$.

We may assume that the system has the property of microscopic reversibility
(i.e., of detailed
balancing) according to which the probability of the transition
$b^{\prime}\to b$ is equal to the
probability of the time-reversed transition $\hat{b}\to\hat{b^{\prime}}$:
$P_{bb^{\prime}}=P_{\hat{b^{\prime}}\hat b}$, as expected for instance in
absence of a magnetic
field.  As a consequence of detailed balancing, the matrix $\mathsf P$ is a
bi-stochastic
matrix, i.e., it satisfies $\sum_{b}P_{bb^{\prime }}=\sum_{b^{\prime
}}P_{bb^{\prime }}=1$,
whereupon the vector $\lbrace 1,1,\ldots,1 \rbrace$ is both a right and
left eigenvector of
$\mathsf P$ with eigenvalue $1$.  This is the case for a finite graph with
transition
probabilities $P_{bb^{\prime }}=\vert T_{bb^{\prime }}\vert^2$ given by the
amplitudes (\ref{Ts}).

\subsection{The classical dynamics on graphs as a random suspended flow}

\label{subsec.susp}

The description given above is analogous to the dynamics of a so-called
suspended flow \cite{GaspEigenDiff}. In fact, we can consider the set of
points $\left\{
x_{b}=0, \ \forall b\right\}$, i.e., the set of
all vertices, as a surface of section.  We attach to each of these points a
segment
(here the directed bond) characterized by a coordinate $0<x <l_{b}$. When the
trajectory reaches the point $x=l_{b}$ it performs another passage through the
surface. Thus the flow is suspended over the Poincar\'{e} surface of section
made of the vertices in the phase space of the directed bonds.

For convenience, instead of the previous notation $x_{b}$, the position
(in phase space) will be referred to as the pair $\left[ b,x \right]$ where
$b$ indicates the directed bond and $x$ is the position on this bond,
i.e., $0<x <l_{b}$.

A realization of the random process on the graph (i.e., a trajectory) can
be identified with the sequence of traversed bonds $\cdots
b_{-2}b_{-1}b_{0}b_{1}b_{2}\cdots $ (which is enough to determine the
evolution on the surface of section). The probability of such a trajectory
is given by $\cdots
P_{b_{2}b_{1}}P_{b_{1}b_{0}}P_{b_{0}b_{-1}}P_{b_{-1}b_{-2}}\cdots $.

An initial condition $[b_0,x]$ of this trajectory is denoted by the dotted
bi-infinite sequence
$\cdots b_{-2}b_{-1}\cdot b_{0}b_{1}b_{2}\cdots $ together with the
position $0\leq x <
l_{b_0}$.  For a given trajectory, we divide the time axis in intervals of
duration $\frac{
l_{b_{n}}}{v}$ extending from
\begin{equation*}
\frac{l_{b_{0}}-x}{v}+\frac{l_{b_{1}}}{v}+\cdots
+\frac{l_{b_{n-2}}}{v}+\frac{ l_{b_{n-1}}}{v}\text{\quad to\quad
}\frac{l_{b_{0}}-x}{v}+\frac{l_{b_{1}}}{v}+\cdots +\frac{
l_{b_{n-2}}}{v}+\frac{l_{b_{n-1}}}{v}+\frac{l_{b_{n}}}{v}
\end{equation*}
where $v$ is the velocity of the particle which travels freely in the bonds.
At each vertex, the particle changes its direction but keeps constant its
kinetic energy.

For a trajectory $p$ that, at time $t=0$, is at the position $[b_{0},x ]$
we define the forward evolution operator $\Phi_{p}^{t}$ with $t>0$ by
\begin{equation}
\Phi _{p}^{t}\left[ b_{0},x \right] =\left[ b_{0},vt+x \right] \qquad
\text{if } \qquad 0<x +vt<l_{b_{0}}  \label{Evol.bond}
\end{equation}
i.e., the evolution is the one of
a free particle as long as the particle stays in the bond $b_0$, and
\begin{multline}
\Phi _{p}^{t}\left[ b_{0},x \right] =\left[ b_{n},x
+vt-l_{b_{n-1}}-l_{b_{n-2}}-\cdots -l_{b_{0}}\right] \\
\text{if\quad }0<x +vt-l_{b_{n-1}}-l_{b_{n-2}}-\cdots -l_{b_{0}}<l_{b_{n}}
\label{Evol.bond2}
\end{multline}
which follows from the fact that, for the given trajectory $p$, the bond
and the position
where the particle stands at a given time is fixed by the lengths
traversed at previous times and by the constant velocity $v$. Analogously,
we also introduce a
backward evolution for $t<0$:
\begin{equation*}
\Phi _{p}^{-|t|}\left[ b_{0},x \right] =\left[ b_{0},x -v|t|\right]
\qquad \text{if }\qquad 0<x -v|t|<l_{b_{0}}
\end{equation*}
and
\begin{multline*}
\Phi _{p}^{-|t|}\left[ b_{0},x \right] =\left[ b_{-n},x
-v|t|+l_{b_{-n}}+l_{b_{-n+1}}+\cdots +l_{b_{-1}}\right] \\
\text{if\quad }0<x -v|t|+l_{b_{-n}}+l_{b_{-n+1}}+\cdots
+l_{b_{-1}}<l_{b_{-n}}
\end{multline*}

\section{The Liouvillian description}

\label{sec.liouville}

\subsection{The Frobenius-Perron operator}

\label{subsec.FP}

On the graph, we want to study the time evolution of the probability
density $\rho
\left( \left[ b,x \right] ,t\right) $.  This density determines the
probability
$\rho \left( \left[ b,x \right] ,t\right) dx $ of finding the particle in
the bond $b$ with position in $[x ,x +dx ]$ at time $t$.

For a general Markov process, the time evolution of the probability density
is ruled by the
Chapman-Kolmogorov equation
\begin{equation}
\rho (\xi,t)=\int d\xi_{0}\,{\cal P}\left( \xi,t|\xi_{0},t_{0}\right) \rho
(\xi_{0},t_{0})
\label{Chap.Kol}
\end{equation}
where ${\cal P}\left( \xi,t|\xi_{0},t_{0}\right)$ is the conditional
probability density that the
particle be in the state $\xi$ at time $t$ given that it was in the state
$\xi_0$ at the initial time $t_0$.  This conditional probability density
defines the integral
kernel of the time-evolution operator, which is linear.  The conditional
probability density can
be expressed as a sum (or integral) over all the paths joining the initial
state to the final one
within the given lapse of time.

In the case of graphs, each state is given by an directed bond and a
position on this bond:
$\xi=[b,x]$.  A path or trajectory is a bi-infinite sequence of directed
bonds as described in
the preceding section.  As soon as the path or trajectory $p$ is known, the
sequence of visited
bonds is fixed so that the motion is determined to be the time translation
at velocity $v$ given
by Eqs. (\ref{Evol.bond})-(\ref{Evol.bond2}).  In this case, the
conditional probability density
of finding the particle in position $[b,x]$ at time $t$ given it was in
$[b_0,x_0]$ at the
initial time $t_0=0$ is given by a kind of Dirac delta density $\delta
\left( \left[ b,x \right] -
\Phi_{p}^{t}\left[ b_{0},x _{0}\right] \right)$.  Along the path $p$, the
particle meets several
successive vertices where the conditional probability density is expressed
in terms of the
conditional probability to reach the final bond $b=b_n$ within the time
$t$, given the initial
condition $[b_0,x_0]$:
\begin{equation}
P_p(t,[b_0,x_0]) \equiv ({\mathsf P}^n)_{bb_0}=P_{b
b_{n-1}}P_{b_{n-1}b_{n-2}}\cdots
P_{b_{2}b_{1}}P_{b_{1}b_{0}}
\label{Prob.path}
\end{equation}
We notice that the integer $n$ is fixed by the trajectory $p$, the initial
condition $[b_0,x_0]$,
and the elapsed time $t$.  The number of the path probabilities
(\ref{Prob.path}) which are
non-vanishing is always finite if there is no sequence of lengths
accumulating to zero for the
graph under study.

For a graph, the kernel of the evolution operator is thus given by
\begin{equation}
{\cal P}\left( \left[ b,x \right] ,t|\left[ b_{0},x _{0}\right] ,0\right)
=\sum_{\{p\}}P_p(t,[b_0,x_0]) \; \delta \left( \left[ b,x \right] -\Phi
_{p}^{t}\left[
b_{0},x _{0}\right] \right) ,
\label{kernel}
\end{equation}
where the sum is performed over a finite number of paths.
By analogy with Eq. (\ref{Chap.Kol}) the density is given by
\begin{equation*}
\rho \left( \left[ b,x \right] ,t\right)
=\sum_{b_{0}}\int_{0}^{l_{b_{0}}}dx _{0}\left(
\sum_{\{p\}}P_p(t,[b_0,x_0])\; \delta
\left( \left[ b,x \right] -\Phi _{p}^{t}\left[ b_{0},x _{0}\right]
\right) \right) \rho \left( \left[ b_{0},x _{0}\right] ,0\right)
\end{equation*}
and integrating the Dirac delta density\footnote{
We emphasize the analogy with deterministic
processes for which ${\cal P}(x,t|x_{0},0)=\delta \left[x-\phi
^{t}(x_{0})\right]$ so that $
\rho (x,t)=\left| \frac{\partial \phi ^{-t}}{\partial x}\right| \rho \left[\phi
^{-t}(x),0\right]$.} we finally get
\begin{equation}
\rho \left( \left[ b,x \right] ,t\right) =\sum_{\{p\}}P_p(t,\Phi
_{p}^{-t}\left[ b,x \right])
\ \rho \left( \Phi _{p}^{-t}\left[ b,x \right] ,0\right) .  \label{Ev.dens}
\end{equation}
where the sum is over all the trajectories that go backward in time from
the current point
$\left[b,x \right] $.

In this way, we have defined the Frobenius-Perron operator $\hat{P}^{t}$ as
\begin{equation}
\hat{P}^{t}\,F\left[ b,x \right] =\sum_{\{p\}}P_p(t,\Phi_{p}^{-t}\left[ b,x
\right]) \
F\left(\Phi _{p}^{-t}\left[ b,x \right] \right)  \label{Frob-Perron}
\end{equation}
where $\left\lbrace F[b,x]\right\rbrace$ is a vector of $2B$
functions defined on the directed bonds.

We now turn to the determination of the spectrum of the Frobenius-Perron
operator. With this
aim, we take the Laplace transform of the
Frobenius-Perron operator given by Eq. (\ref{Frob-Perron}):
\begin{equation*}
\int_{0}^{\infty }\mbox{e}^{-st}\hat{P}^{t}\, F\left[ b,x \right]
dt=\int_{0}^{\infty }dt\,\mbox{e}^{-st}\sum_{\{p\}}P_p(t,\Phi
_{p}^{-t}\left[ b,x \right]) \ F\left(\Phi
_{p}^{-t}\left[ b,x \right] \right)
\end{equation*}
In order to evaluate the Laplace transform, we have to decompose the sum
over the paths $\{ p\}$
into the different classes of terms corresponding to paths in which
$n=0,1,2,...$
bonds are visited during the time $t$.  This decomposition leads to
\begin{multline*}
\sum_{\{p\}}\int_{0}^{\infty }\mbox{e}^{-st}P_p \ F\left( \Phi
_{p}^{-t}\left[ b,x \right]
\right) dt=\int_{0}^{\frac{x }{v}}\mbox{e}^{-st} F\left[ b,x -vt\right] dt
\\
+\sum_{n=1}^{\infty }\sum_{\{p^{-n}\}}\int_{\frac{x }{v}+\Sigma
_{i=1}^{n-1}\frac{
l_{b_{-i}}}{v}}^{\frac{x }{v}+\Sigma _{i=1}^{n}\frac{l_{b_{-i}}}{v}
}\,\mbox{e}^{-st}\,P_{p^{-n}}\,F\left[ b_{-n},x -vt+\sum_{i=1}^{n}l_{b_{-i}}
\right] dt
\end{multline*}
with $P_{p^{-n}}=P_{bb_{-1}}P_{b_{-1}b_{-2}}\cdots P_{b_{-n+1}b_{-n}}$
the probability of a path $p^{-n}$ and $\sum_{\{p^{-n}\}}$ the sum over these
trajectories.
 A change of variable transforms the previous
equation in the following
\begin{multline}
\sum_{\{p\}}\int_{0}^{\infty }\mbox{e}^{-st}P_p\; F\left( \Phi
_{p}^{-t}\left[ b,x \right]
\right) =\frac{1}{v}\; \mbox{e}^{-s\frac{x }{v}}\left\lbrace \int_{0}^{x
}\mbox{e}^{s\frac{
x ^{\prime }}{v}}F\left[ b,x ^{\prime }\right] dx ^{\prime } \right.\\ \left.
+\sum_{n=1}^{\infty }\sum_{\{p^{-n}\}}\left(
\prod_{i=1}^{n}Q_{b_{-i+1}b_{-i}}\right)
\int_{0}^{l_{b_{-n}}}\mbox{e}^{-s\frac{x ^{\prime }}{v}}F\left[ b_{-n},x
^{\prime }\right] dx ^{\prime }\right\rbrace
\label{neweq}
\end{multline}
where we introduced the quantity
\begin{equation}
Q_{bb^{\prime }}(s)=P_{bb^{\prime }}\; \mbox{e}^{-s\frac{l_{b^{\prime }}}{v}}
\label{Eq.Q}
\end{equation}
and here we identify $b_{0}$ with $b$. Now we have to perform the sum over all
the realizations in the right-hand side of Eq. (\ref{neweq}).
This is a sum over all the trajectories $
b_{-1}b_{-2}\cdots b_{-n+1}b_{-n}$ which leads to the formation\footnote{$
\sum_{b_{-1}\cdots b_{-n}}Q_{b_{0}b_{-1}}\cdots
Q_{b_{-n+1}b_{-n}}=\sum_{b_{-n}}({\mathsf Q}^{n})_{b_{0}b_{-n}}$} of the
matrix ${\mathsf
Q}^{n}$, i.e., $\mathsf Q$ raised to the power $n$.  Accordingly, we get
\begin{multline}
\int_{0}^{\infty }\mbox{e}^{-st}\hat{P}^{t}\,F\left[ b,x \right] dt
=\frac{1}{v}\;
\mbox{e}^{-sx/v}\left\lbrace \int_{0}^{x }\mbox{e}^{s\frac{x ^{\prime
}}{v}}F\left[ b,x ^{\prime }\right] dx
^{\prime }  \right. \\\left.  +
\sum_{n=1}^{\infty }\sum_{b_{-n}}\left( {\mathsf Q}^{n}\right)_{bb_{-n}}
\int_{0}^{l_{b_{-n}}}\mbox{e}^{s\frac{x ^{\prime }}{v}}F\left[ b_{-n},x
^{\prime }\right] dx ^{\prime }\right\rbrace.  \label{lap.trans}
\end{multline}

We define the vector ${\mathbf f}(s)=\lbrace f_1(s),...,f_{2B}(s)\rbrace$
with the components of
this vector given by the functions
\begin{equation}
f_{b}(s) =\int_{0}^{l_{b}}\mbox{e}^{s\frac{x ^{\prime
}}{v}}F\left[b,x^{\prime }\right] dx ^{\prime }
\label{fs}
\end{equation}
The matrix ${\mathsf Q}(s)$ acts on these vectors ${\mathbf f}(s)$ through
the relation
\begin{equation}
\left({\mathsf Q}\cdot{\mathbf f}\right)_b(s) =\sum_{b^{\prime }}Q_{bb^{\prime
}}(s) f_{b^{\prime}}(s) .  \label{op.R}
\end{equation}

This matrix can be interpreted as the Frobenius-Perron operator of the
evolution reduced to the surface of section \cite{GaspEigenDiff}. This
matrix depends on the
Laplace variable $s$ which will give the relaxation rate of the system.
With these definitions,
we can write
\begin{equation*}
\sum_{n=1}^{\infty }\sum_{b_{-n}}\left( {\mathsf Q}^{n}\right)_{bb_{-n}}
\int_{0}^{l_{b_{-n}}}\mbox{e}^{s\frac{x ^{\prime }}{v}}F\left[ b_{-n},x
^{\prime }\right] dx ^{\prime }
= \sum_{n=1}^{\infty }\left({\mathsf Q}^{n}\cdot{\mathbf f}\right)_b(s)
= \left(\frac{\mathsf Q}{{\mathsf I}-{\mathsf Q}}\cdot{\mathbf f}\right)_b(s)
\end{equation*}
where we used the relation $\sum_{n=1}^{\infty }{\mathsf
Q}^{n}=\frac{{\mathsf Q}}{{\mathsf
I}-{\mathsf Q}}$.  As a consequence, Eq. (\ref{lap.trans}) becomes
\begin{equation}
\int_{0}^{\infty }\mbox{e}^{-st}\hat{P}^{t}\,F\left[ b,x \right] dt=\frac{1}{
v}\; \mbox{e}^{-sx /v}\left\{ \int_{0}^{x }\mbox{e}^{s\frac{x ^{\prime
}}{v}}F\left[
b,x ^{\prime }\right] dx ^{\prime }+ \left(\frac{\mathsf Q}{{\mathsf
I}-{\mathsf Q}}\cdot{\mathbf
f}\right)_b(s) \right\} .  \label{Lap.Evol}
\end{equation}

We are now at a few steps from determining the eigenvalues (and
eigenvectors) of $\hat{P}^{t}$. This is done by first studying the
solutions of
\begin{equation*}
{\mathsf Q}(s)\cdot{\mathbf f}(s) ={\mathbf f}(s) .
\end{equation*}
These solutions exist only if $s$ belongs to the (complex) set $\left\{
s_{j}\right\} $ of
solutions of the following characteristic determinant
\begin{equation}
\det \left[ {\mathsf I}-{\mathsf Q}(s)\right] =0  \label{detI-Q}
\end{equation}
We denote these particular vectors by $\pmb{\chi}_{j}$ and their components
by $\chi_j[b]$
with $b=1,...,2B$, whereupon
\begin{equation}
{\mathsf Q}(s_j)\cdot\pmb{\chi}_j =\pmb{\chi}_j .  \label{eigenR}
\end{equation}
The left vector which is adjoint to the right vector $\pmb{\chi}_{j}$ is
given by
\begin{equation}
{\mathsf Q}(s_j)^{\dagger}\cdot\tilde{\pmb{\chi}}_j =\tilde{\pmb{\chi}}_j .
\label{adjeigenR}
\end{equation}

The relation to the eigenvalue problem of the flow is established as
follows. Suppose that $\Psi _{j}\left[ b,x \right] $ is an eigenstate
of $\hat{P}^{t}$ with eigenvalue $\mbox{e}^{s_{j}t}$ [for the moment $s_{j}$
is not determined but we call it this way because it will turn out
to be one of the solutions of
Eq. (\ref{detI-Q})], i.e.,
\begin{equation}
\hat{P}^{t}\,\Psi _{j}\left[ b,x \right] =\mbox{e}^{s_{j}t}\Psi _{j}\left[
b,x \right]  \label{Eigen.Flow}
\end{equation}
with ${\mathrm Re}\,s_{j}\leq 0$ because the density is not expected to
increase with the time. For the forward semigroup, the zeros of Eq. (\ref
{detI-Q}) are expected in the region ${\mathrm Re}\,s_{j}\leq 0$.

Taking the Laplace transform of Eq. (\ref{Eigen.Flow}), we get
\begin{equation*}
\int_{0}^{\infty }dt\,\mbox{e}^{-st}\,\hat{P}^{t}\,\Psi _{j}\left[ b,x
\right] =\int_{0}^{\infty }dt\,\mbox{e}^{(s_{j}-s)t}\Psi _{j}\left[ b,x
\right] =
\frac{\Psi _{j}\left[ b,x \right] }{s-s_{j}}
\end{equation*}
If we introduce the vector $\pmb{\Upsilon}(s)=\lbrace
\Upsilon_1(s),...,\Upsilon_{2B}(s)\rbrace$ defined by the components
\begin{equation*}
\Upsilon_{b}(s) =\int_{0}^{l_{b}}dx \, \mbox{e}^{s\frac{x }{v}
}\Psi _{j}\left[ b,x \right]
\end{equation*}
and use the same calculation that led to Eq. (\ref{Lap.Evol}), the
eigenvalue equation becomes
\begin{equation}
\int_{0}^{x }dx ^{\prime }\,\mbox{e}^{s\frac{x ^{\prime }}{v}}\Psi _{j}
\left[ b,x ^{\prime }\right] +\left(\frac{{\mathsf Q}}{{\mathsf I}-{\mathsf
Q}}\cdot
\pmb{\Upsilon}\right)_b(s) =\frac{v\; \mbox{e}^{s\frac{x }{v}}\,\Psi
_{j}\left[ b,x \right] }{
s-s_{j}}.  \label{Eig.Sol}
\end{equation}
for $0<x <l_{b}$.
Setting $x =0$ in Eq. (\ref{Eig.Sol}), we have
\begin{equation*}
(s-s_{j})\; {\mathsf Q}(s)\cdot\pmb{\Upsilon}(s) =v\; \left[{\mathsf
I}-{\mathsf
Q}(s)\right]\cdot\pmb{\Psi}_{j}
\end{equation*}
with the vector $\pmb{\Psi}_j=\lbrace\Psi_j[b,0]\rbrace_{b=1}^{2B}$.
For $s=s_{j}$, we get that
\begin{equation*}
{\mathsf Q}(s_j)\cdot\pmb{\Psi}_{j} = \pmb{\Psi}_{j}
\end{equation*}
which shows that $s_{j}$ is a solution of Eq. (\ref{detI-Q}) as we
anticipated and that the eigenstate of the flow at $x =0$, $\Psi _{j}
\left[ b,0\right]$, may be identified with the vector which is solution of
Eq. (\ref{eigenR}):
\begin{equation*}
\Psi_j[b,0] = \chi _{j}\left[ b\right] =\sum_{b^{\prime
}}Q_{bb^{\prime }}(s_j) \; \chi _{j}\left[ b^{\prime }\right]
\end{equation*}
To determine the eigenstates of the flow for the other values of $x $ we
differentiate Eq. (\ref{Eig.Sol}) with respect to $x $ and we get
\begin{equation*}
\mbox{e}^{s\frac{x }{v}}\,\Psi _{j}\left[ b,x \right]
=\frac{s\,\mbox{e}^{s\frac{
x }{v}}\,\Psi _{j}\left[ b,x \right] }{s-s_{j}}+\frac{v\,\mbox{e}^{s\frac{
x }{v}}\,\partial _{x }\Psi _{j}\left[ b,x \right] }{s-s_{j}}
\end{equation*}
from which we obtain
\begin{equation*}
\partial _{x }\Psi _{j}\left[ b,x \right] =-\frac{s_{j}}{v}\Psi _{j}
\left[ b,x \right]
\end{equation*}
the integration of which gives
\begin{equation}
\Psi _{j}\left[ b,x \right] =\mbox{e}^{-s_{j}\frac{x }{v}}\Psi _{j}\left[ b,0
\right] =\mbox{e}^{-s_{j}\frac{x }{v}}\chi _{j}\left[ b\right] \quad
\text{for }
0\leq x \leq l_{b}. \label{right.eigen}
\end{equation}
The eigenstate increases exponentially along each directed bond.  This
exponential increase
does not constitute a problem because the time
evolution generates the overall exponential decay of Eq.
(\ref{Eigen.Flow}). Therefore, we see
that the vectors $\chi _{j}\left[ b\right] $ which are solutions of Eq.
(\ref{eigenR}) determine
the eigenstates of the Frobenius-Perron operator. These eigenstates are
very important in
nonequilibrium statistical mechanics since they provide the link between the
microscopic and the phenomenological description of the system
\cite{GaspEigenDiff}.

\subsection{A master-equation approach}

\label{subsec.master}

In this section, we develop an alternative derivation of the results of
the previous subsection but here by using a master equation.

If at a given time $t$ the particle is at the end of a bond, say $\left[
b^{\prime },l_{b^{\prime }}\right] $, the particle has to go instantaneously
to another directed bond with probability $P_{bb^{\prime }}$, i.e.,
\begin{equation}
\rho \left( \left[ b,0\right] ,t\right) =\sum_{b^{\prime }}P_{bb^{\prime
}}\rho \left( \left[ b^{\prime },l_{b^{\prime }}\right] ,t\right)
\label{Ev.0}
\end{equation}
Now, since the evolution is deterministic along the bonds [see Eq. (\ref
{Evol.bond})] we have that\footnote{
This is because $\rho \left( \left[ b,x \right] ,t+\frac{x }{v}\right)
=\hat{P}^{\frac{x }{v}}\rho \left( \left[ b,x \right] ,t\right)
=\rho \left( \Phi ^{-\frac{x }{v}}\left[ b,x \right] ,t\right) =\rho
\left( \left[ b,0\right] ,t\right)$.}
\begin{equation*}
\rho \left( \left[ b,x \right] ,t+\frac{x }{v}\right) =\rho \left(
\left[ b,0\right] ,t\right)
\end{equation*}
and also
\begin{equation*}
\rho \left( \left[ b^{\prime },l_{b^{\prime }}\right] ,t\right) =\rho \left(
\left[ b^{\prime },l_{b^{\prime }}-vt^{\prime }\right] ,t-t^{\prime }\right).
\end{equation*}
Choosing $t^{\prime }=\frac{l_{b^{\prime }}-x ^{\prime }}{v}$, we have
\begin{equation*}
\rho \left( \left[ b^{\prime },l_{b^{\prime }}\right] ,t\right) =\rho \left(
\left[ b^{\prime },x ^{\prime }\right] ,t-\frac{l_{b^{\prime }}-x^{\prime
}}{v}\right)
\end{equation*}
where $x^{\prime}$ is arbitrary.
Hence, Eq. (\ref{Ev.0}) becomes
\begin{equation*}
\rho \left( \left[ b,x \right] ,t+\frac{x }{v}\right) =\sum_{b^{\prime
}}P_{bb^{\prime }}\rho \left( \left[ b^{\prime },x_{b^{\prime}}\right] ,t-
\frac{l_{b^{\prime }}-x_{b^{\prime}}}{v}\right)
\end{equation*}
where $x_{b^{\prime}}$ may be chosen arbitrarily on each bond ${b^{\prime}}$.
With the replacement $t+\frac{x }{v}\rightarrow t$, we finally obtain
\begin{equation}
\rho \left( \left[ b,x \right] ,t\right) =\sum_{b^{\prime }}P_{bb^{\prime
}}\rho \left( \left[ b^{\prime },x_{b^{\prime}}\right]
,t-\frac{x+l_{b^{\prime }}
-x_{b^{\prime}} }{v}\right)  \label{Ev.t}
\end{equation}
This is the master equation which rules the time evolution on the graph. It
is a Markovian equation with a time delay. The master equation (\ref{Ev.t})
differs from Eq. (\ref{Ev.dens}) in the sense that Eq. (\ref{Ev.t}) relates
the probability densities before and after the transitions although Eq. (\ref
{Ev.dens}) relates the density at time $t$ to the initial density through a
varying number of transitions, depending on the path $p$.

Stationary solutions of the equation (\ref{Ev.t}), satisfying $\rho \left(
\left[
b,x \right] ,t\right) =\widetilde{\rho }\left( \left[ b,x \right]
\right)$ for all $t$, exist if the matrix ${\mathsf P}$ as an eigenvalue
$1$. This
is the case for closed graphs. For open graphs, the density decays in time in
a way that we shall determine below.

The master equation (\ref{Ev.t}) can be iterated. For instance the second
iteration gives
\begin{equation*}
\rho \left( \left[ b,x \right] ,t\right) =\sum_{b^{\prime }b^{\prime
\prime }}P_{bb^{\prime }}P_{b^{\prime }b^{\prime \prime }}\; \rho \left( \left[
b^{^{\prime \prime }},x_{b^{\prime}b^{\prime \prime }}\right] ,t-\frac{x+
l_{b^{\prime }}+l_{b^{\prime \prime }}-
x_{b^{\prime}b^{\prime \prime }} }{v}\right)
\end{equation*}
and in general
\begin{align}
\rho \left( \left[ b,x \right] ,t\right) & =\sum_{b^{\prime }b^{\prime
\prime }\cdots b^{(n)}}P_{bb^{\prime }}P_{b^{\prime }b^{\prime \prime
}}\cdots P_{b^{(n-1)}b^{(n)}}  \notag \\
& \times \ \rho \left( \left[ b^{^{(n)}},x_{b^{\prime}\cdots
b^{(n)}}\right] ,t-\frac{x+
\sum_{i=1}^{n}l_{b^{(i)}}-x_{b^{\prime}\cdots b^{(n)}}}{v}\right) .
\label{Ev.Iter}
\end{align}

There exists an integer $n$ for which we find (at least one) solution of
\begin{equation}
t-\frac{x+l_{b^{\prime \prime }}+\cdots
+l_{b^{(n)}}-x_{b^{\prime}\cdots b^{(n)}} }{v}=0\quad \text{with}\quad
0<x_{b^{\prime}\cdots b^{(n)}}<l_{b^{(n)}}
\label{Cond.Inc}
\end{equation}
for some path $b^{\prime }b^{\prime \prime }\cdots b^{(n)}$. Accordingly,
we split the
sum (\ref{Ev.Iter}) in two terms, the first one with all the possible paths
for which there exists a value $x_{b^{\prime}\cdots b^{(n)}}$ which solves
Eq. (\ref{Cond.Inc})
with the smallest integer $n$ ($\sum '$ denotes the sum over these paths)
and the other term containing the rest of (\ref{Ev.Iter}),
i.e.,
\begin{multline*}
\rho \left( \left[ b,x \right] ,t\right) =\sum '_{b^{\prime }b^{\prime
\prime }\cdots b^{(n)}}P_{bb^{\prime }}P_{b^{\prime }b^{\prime
\prime }}\cdots P_{b^{(n-1)}b^{(n)}}\rho \left( \left[
b^{^{(n)}},x_{b^{\prime}\cdots b^{(n)}}\right] ,0\right)  \\
+ \sum_{b^{\prime }b^{\prime \prime }\cdots b^{(n)}}P_{bb^{\prime
}}P_{b^{\prime }b^{\prime \prime }}\cdots P_{b^{(n-1)}b^{(n)}}\rho \left(
\left[ b^{^{(n)}},x_{b^{\prime}\cdots b^{(n)}}\right] ,t-\frac{
x+\sum_{i=1}^{n}l_{b^{(i)}}-x_{b^{\prime}\cdots b^{(n)}} }{v}\right)
\end{multline*}
and we proceed iteratively with the second term, that is, we look for the
smallest $n$ for which there exists a path for which a solution of Eq. (\ref
{Cond.Inc}) exists and so on. Thus we finally have
\begin{equation}
\rho \left( \left[ b,x \right] ,t\right) =\sum_{n}\sum_{b^{\prime
}b^{\prime \prime }\cdots b^{(n)}}P_{bb^{\prime }}P_{b^{\prime }b^{\prime
\prime }}\cdots P_{b^{(n-1)}b^{(n)}}\rho \left( \left[
b^{^{(n)}},x_{b^{\prime}\cdots b^{(n)}}
\right] ,0\right)  \label{Eq.funda}
\end{equation}
with
\begin{equation*}
x_{b^{\prime}\cdots b^{(n)}}=x -vt+\sum_{i=1}^{n}l_{b^{(i)}}.
\end{equation*}
Accordingly, $\rho \left( \left[ b,x \right] ,t\right)$ is given by a sum
over the initial conditions $\left[ b^{^{(n)}},x_{b^{\prime}\cdots
b^{(n)}}\right] $\ and
over all the paths that connect $\left[ b^{^{(n)}},x_{b^{\prime}\cdots
b^{(n)}}\right] $
with $\left[ b,x \right] $ in a time $t$. Each given path contributes
to this sum by its probability
multiplied by the probability density $\rho \left( \left[
b^{^{(n)}},x_{b^{\prime}\cdots b^{(n)}}\right] ,0\right) $. Using the
notation introduced
before [see Eq. (\ref{Prob.path})] Eq. (\ref{Eq.funda}) can be written as
\begin{equation}
\rho \left( \left[ b,x \right] ,t\right)
=\sum_{\{p\}}P_p\left(t,\Phi_p^{-t}[b,x]\right)\; \rho
\left( \left[ b^{^{(n)}},x_{b^{\prime}\cdots b^{(n)}}\right] ,0\right)
\label{Ev.dens2}
\end{equation}
with $x_{b^{\prime}\cdots b^{(n)}}=x -vt+\sum_{i=1}^{n}l_{b^{(i)}}$. We see
that this
equation coincides with Eq. (\ref{Ev.dens}), which shows that both approaches
are equivalent. In fact, if we write the sequence $bb^{\prime }\cdots b^{(n)}$
in the form $bb_{-1}\cdots b_{-n}$ and if we remember that $\Phi _{p}^{-t}
\left[ b,x \right] =\left[ b_{-n},x -vt+\sum_{i=1}^{n}l_{b_{-i}}\right]
$ if $0<x -vt+\sum_{i=1}^{n}l_{b_{-i}}<l_{b_{-n}}$, the equivalence is
established.

Now we turn to the determination of the spectrum of the Frobenius-Perron
operator from the master equation. Taking the Laplace transform of Eq. (\ref
{Ev.t}) or equivalently of
\begin{equation*}
\rho \left( \left[ b,0\right] ,t-\frac{x }{v}\right) =\sum_{b^{\prime
}}P_{bb^{\prime }}\rho \left( \left[ b^{\prime },0\right] ,t-\frac{
l_{b^{\prime }}+x }{v}\right)
\end{equation*}
we have
\begin{equation}
\int_{-\frac{x }{v}}^{\infty }\mbox{e}^{-st^{\prime }}\rho \left( \left[ b,0
\right] ,t^{\prime }\right) dt^{\prime }=\sum_{b^{\prime }}P_{bb^{\prime
}}\,\mbox{e}^{-s\frac{l_{b^{\prime }}}{v}}\int_{-\frac{x
}{v}-\frac{l_{b^{\prime
}}}{v}}^{\infty }\mbox{e}^{-st^{\prime }}\rho \left( \left[ b^{\prime
},0\right]
,t^{\prime }\right) dt^{\prime }
\label{truc}
\end{equation}
after some simple changes of variable.\footnote{
At the left-hand side $t^{\prime }=t-x /v$ and at the right-hand side $
t^{\prime }=t-x /v-l_{b^{\prime }}/v.$} Considering the definition
of Eq. (\ref{Eq.Q}) and that $\rho \left(
\left[ b,0\right] ,t<0\right) =0$, Eq. (\ref{truc}) reads
\begin{equation}
\int_{0}^{\infty }\mbox{e}^{-st^{\prime }}\rho \left( \left[ b,0\right]
,t^{\prime
}\right) dt^{\prime }=\sum_{b^{\prime }}Q_{bb^{\prime }}(s)\int_{0}^{\infty
}\mbox{e}^{-st^{\prime }}\rho \left( \left[ b^{\prime },0\right] ,t^{\prime
}\right) dt^{\prime }.  \label{rho.map}
\end{equation}
Defining
\begin{equation*}
\rho_{b}(s) =\int_{0}^{\infty }\mbox{e}^{-st}\rho \left( \left[ b,0
\right] ,t\right) dt
\end{equation*}
Eq. (\ref{rho.map}) becomes
\begin{equation*}
\rho _{b}(s) =\sum_{b^{\prime }}Q_{bb^{\prime }}(s)\rho _{b}(s)
\end{equation*}
which has solutions only if $s$ belongs to the set $\left\{ s_{j}\right\} $
of solutions of
\begin{equation*}
\det \left[ {\mathsf I}-{\mathsf Q}(s)\right] =0
\end{equation*}
and
\begin{equation}
\pmb{\chi}_{j}={\mathsf Q}\left( s_{j}\right)\cdot\pmb{\chi}_{j}\ .
\label{Q.map}
\end{equation}
The eigenstates of the flow
\begin{equation}
\rho_{j}\left( \left[ b,x \right] ,t\right) =\mbox{e}^{s_{j}t}\rho _{j}\left(
\left[ b,x \right] ,0\right)  \label{eigen.eq'}
\end{equation}
are determined as follows. We replace Eq. (\ref{eigen.eq'}) in the master
equation (\ref{Ev.t}) from where we directly get
\begin{equation}
\left\{ \mbox{e}^{s_{j}\frac{x }{v}}\rho _{j}\left( \left[ b,x \right]
,0\right) \right\} =\sum_{b^{\prime }}Q_{bb^{\prime }}\left( s_{j}\right)
\left\{ \mbox{e}^{s_{j}\frac{x }{v}}\rho _{j}\left( \left[ b^{\prime },x
\right] ,0\right) \right\}.  \label{Q.map2}
\end{equation}
Comparing Eq. (\ref{Q.map}) and Eq. (\ref{Q.map2}) we have that the
eigenstates of the Frobenius-Perron operator of the flow are given by
\begin{equation*}
\rho _{j}\left( \left[ b,x \right] ,0\right) =\mbox{e}^{-s_{j}\frac{x }{v}
}\chi _{j}\left[ b\right] \quad \text{for }0<x <l_{b}
\end{equation*}
and we have recovered the same results as previously obtained with the
suspended-flow approach.


\section{The relaxation and ergodic properties}

\label{sec.relax}

\subsection{The spectral decomposition of the Frobenius-Perron operator}

Thanks to the knowledge of Frobenius-Perron operator,
we can study the time evolution of the statistical averages of the physical
observables $A[b,x]$
defined on the bonds of the graphs as
\begin{equation}
\langle A\rangle_t = \sum_{b=1}^{2B} \frac{1}{l_b} \int_0^{l_b} A[b,x] \;
\rho([b,x],t) dx =
\langle A \vert \hat P^t \rho_0 \rangle
\end{equation}
where $\rho_0$ denotes the initial probability density and where we have
introduced the inner
product
\begin{equation}
\langle F \vert G \rangle = \sum_{b=1}^{2B} \frac{1}{l_b} \int_0^{l_b}
F[b,x]^* \; G[b,x] dx
\label{inner}
\end{equation}
between two vectors of $2B$ functions $F[b,x]$ and $G[b,x]$ defined on the
bonds.  We have here
used the fact that the Frobenius-Perron operator rules the time evolution
of all the
statistical averages.  If the observable is equal to the unity, $A=1$, the
conservation of the total
probability imposes the normalization condition $\langle 1\rangle_t=1$,
which is satisfied by the
Frobenius-Perron operator.

If we are interested in the time evolution at long times and especially in
the relaxation, we
may consider an asymptotic expansion valid for $t\to +\infty$ of the form
\begin{equation}
\langle A\rangle_t = \langle A \vert \hat P^t \rho_0 \rangle = \sum_j
\langle A \vert
\Psi_j\rangle \ \mbox{e}^{s_jt} \ \langle
\tilde{\Psi}_j\vert \rho_0 \rangle + \cdots
\label{spectral}
\end{equation}
as a sum of exponential functions, together with possible extra terms such
as powers of the time
multiplied by exponentials $t^m\exp(s_jt)$.  In this spectral
decomposition, we have introduced
the right and left eigenstates of the Frobenius-Perron operator
\begin{eqnarray}
\hat P^t \Psi_j = \mbox{e}^{s_jt} \Psi_j \\
\hat P^{t\dagger} \tilde{\Psi}_j = \mbox{e}^{s_j^{*}t} \tilde{\Psi}_j
\end{eqnarray}
Since the Frobenius-Perron operator is not a unitary operator we should
expect Jordan-block
structures and associated root states different from the eigenstates.  Such
Jordan-block
structures are known to generate time dependences of the form
$t^m\exp(s_jt)$.  We shall argue
below that such time behavior is not typical in classical graphs.

With the aim of determining the spectral decomposition (\ref{spectral}), we
take its Laplace
transform:
\begin{equation*}
\int_{0}^{\infty }\mbox{e}^{-st}\langle A\rangle_t
dt= \sum_j \langle A \vert \Psi_j\rangle \ \frac{1}{s-s_j} \ \langle
\tilde{\Psi}_j\vert \rho_0 \rangle + \cdots
\end{equation*}
which allows us to identify the relaxation rates $-s_j$ with the poles of
the Laplace transform and
the eigenstates from the residues of these poles.  For
this purpose, we use the Laplace transform of the Frobenius-Perron operator
given by Eq.
(\ref{Frob-Perron}), which we integrate with the observable quantity
$A[b,x]$.  We get
\begin{eqnarray}
\int_{0}^{\infty }\mbox{e}^{-st}\langle A\vert \hat{P}^{t}\rho_0\rangle \;
dt &=& \sum_b
\frac{1}{ vl_b}\; \int_{0}^{l_b}dx\int_{0}^{x }dx^{\prime}
\mbox{e}^{s\frac{x^{\prime}-x}{v}}
A[b,x] \; \rho_0\left[ b,x^{\prime }\right] \nonumber\\ &+& {\bf a}(s)^{\rm
T}\cdot
\frac{{\mathsf Q}(s)}{{\mathsf I}-{\mathsf Q}(s)}\cdot{\mathbf f}(s)
\label{Lap.Aver}
\end{eqnarray}
where we introduced the vector ${\bf a}(s)$ of components
\begin{equation}
a_{b}(s) =\frac{1}{vl_b}\int_{0}^{l_{b}}\mbox{e}^{-s\frac{x}{v}}A\left[b,x
\right] dx
\label{as}
\end{equation}
and where we used the definition (\ref{fs}) with the initial probability
density $F=\rho_0$.

In Eq. (\ref{Lap.Aver}), the first term is analytic in the complex variable
$s$ and only the
second term can create poles at the complex values $s=s_j$ where the condition
(\ref{detI-Q}) is satisfied.  We suppose here that these poles are simple.
Near the pole
$s=s_j$, we find a divergence of the form
\begin{equation}
\frac{{\mathsf Q}(s)}{{\mathsf I}-{\mathsf Q}(s)} \simeq - \frac{1}{s-s_j} \;
\frac{{\pmb{\chi}_j}\, \tilde{\pmb{\chi}}_j^{\dagger} }
{\tilde{\pmb{\chi}}_j^{\dagger}\cdot
\partial_s{\mathsf Q}(s_j) \cdot {\pmb{\chi}}_j }
\end{equation}
Because of the definition (\ref{Eq.Q}), we have that
\begin{equation}
\tilde{\pmb{\chi}}_j^{\dagger}\cdot
\partial_s{\mathsf Q}(s_j) \cdot {\pmb{\chi}}_j = - \frac{1}{v} \; \sum_b
l_b \; \tilde{\chi}_j[b]^*
\; \chi_j[b]
\end{equation}
In this way, we can identify the relaxation rates of the asymptotic time
evolution of the
physical averages with the roots of the characteristic
determinant (\ref{detI-Q}).  We can also identify the right eigenstates as
\begin{equation}
\langle A\vert\Psi_j\rangle = \sum_b \chi_j[b] \; \frac{1}{l_b}
\int_0^{l_b} \mbox{e}^{-s_j
\frac{x}{v}} A[b,x] \; dx
\label{right.eigenstate}
\end{equation}
which is expected by the previous expression (\ref{right.eigen}) for the
right eigenstates,
and the left eigenstates as
\begin{equation}
\langle \tilde{\Psi}_j\vert\rho_0\rangle = \frac{1}{\sum_{b^{\prime\prime}}
l_{b^{\prime\prime}}
\tilde{\chi}_j[{b^{\prime\prime}}]^* \chi_j[{b^{\prime\prime}}]}
\; \sum_b \tilde{\chi}_j[{b^{\prime}}]^* \; \int_0^{l_{b^{\prime}}}
\mbox{e}^{s_j
\frac{x^{\prime}}{v}} \rho_0[b^{\prime},x^{\prime}] \; dx^{\prime}
\label{left.eigenstate}
\end{equation}
From the definition (\ref{inner}) of the inner product, we infer that the
right eigenstate
associated with the resonance $s_j$ is given by the following vector of
$2B$ functions
\begin{equation}
\Psi_j[b,x] =  \chi_j[b]  \ \mbox{e}^{-s_j\frac{x}{v}}
\label{right.eigenfn}
\end{equation}
while the corresponding left eigenstate is given by
\begin{equation}
\tilde{\Psi}_j[b,x] = \frac{l_b \; \tilde{\chi}_j[b]}
{\sum_{b^{\prime}}l_{b^{\prime}}
\tilde{\chi}_j[{b^{\prime}}] \chi_j[{b^{\prime}}]^*}
\ \mbox{e}^{s_j^*\frac{x}{v}}
\label{left.eigenfn}
\end{equation}
which ends the construction of the spectral decomposition under the
assumption that all the
complex singularities of the Laplace transform of the Frobenius-Perron
operator are isolated
simple poles.

\subsection{The classical zeta function}

The relaxation of the probability density is thus controlled by the relaxation
modes which are given by the eigenvalues and the eigenstates of the
Frobenius-Perron operator.  As we said, the eigenvalues of the
Frobenius-Perron operator are
determined by the solutions $\left\{ s_{j}\right\}$ of the characteristic
determinant
[see Eq. (\ref{detI-Q})]
\begin{equation*}
\det \left[ {\mathsf I}-{\mathsf Q}(s)\right] =0
\end{equation*}
These solutions are complex numbers which are known as the Pollicott-Ruelle
resonances if they are isolated roots.

We will rewrite Eq. (\ref{detI-Q}) in a way that is
reminiscent of the Selberg-Smale zeta function.  With this purpose, we
first consider the
identity
\begin{equation*}
\det ({\mathsf I}-{\mathsf Q})=\exp \; {\rm tr}\; \ln ({\mathsf I}-{\mathsf Q})
\end{equation*}
and the expansion
\begin{equation*}
\ln ({\mathsf I}-{\mathsf Q})=-\sum_{n=1}^{\infty }\frac{{\mathsf Q}^{n}}{n}.
\end{equation*}
from which we infer the following identity
\begin{equation*}
\det ({\mathsf I}-{\mathsf Q})=\exp \left( -\sum_{n=1}^{\infty
}\frac{{\mathrm tr}\;
{\mathsf Q}^{n}}{n}\right)
\end{equation*}
Now, we have that
\begin{equation*}
({\mathsf Q}^{n})_{bb}=\sum_{b_{1}b_{2}\ldots b_{n-1}}Q_{bb_{n-1}}\ldots
Q_{b_{2}b_{1}}Q_{b_{1}b}
\end{equation*}
and from Eq. (\ref{Eq.Q}) we find
\begin{equation*}
{\rm tr}\; {\mathsf Q}^{n}=\sum_{b}({\mathsf
Q}^{n})_{bb}=\sum_{bb_{1}b_{2}\ldots
b_{n-1}}P_{bb_{n-1}}\ldots P_{b_{2}b_{1}}P_{b_{1}b}\;\mbox{e}^{-\frac{s}{v}
(l_{b_{n-1}}+\ldots +l_{b_1}+l_{b})}
\end{equation*}
Note that this is a sum over closed trajectories composed of $n$ lengths in
the graph. The factor
$A_{p}^{2}=P_{bb_{n-1}}\ldots P_{b_{2}b_{1}}P_{b_{1}b}$ plays the role of
the stability factor
of the closed trajectory $bb_1b_2\ldots b_{n-1}$ and following this analogy
we define the Lyapunov
exponent $\lambda_{p}$ per unit time as
\begin{equation}
A_{p}^{2}=\exp (-\lambda _{p} T_p^{(n)})=P_{bb_{n-1}}\ldots
P_{b_{2}b_{1}}P_{b_{1}b}
\label{def.lyap}
\end{equation}
where $T_p^{(n)}$ is the temporal period of this closed trajectory.
We shall consider primitive (or prime) periodic orbits and their repetitions. A
periodic orbit composed of $n$ lengths can be the repetition of a primitive
periodic orbit composed of $n_{p}$ bonds if $n=rn_{p}$ and $r$ is an integer
called the repetition number.  With this definition the total period of the
orbit is given by $T_p^{(n)}=(l_{b_{1}}+l_{b_{2}}+\cdots
+l_{b_{n-1}}+l_{b})/v=rl_{p}/v$ with
$l_{p}$ the length of the primitive periodic orbit.  Accordingly, we have
the following relation
for the Lyapunov exponent of the prime periodic orbit $p=b_1b_2\cdots
b_{n_p}$ composed of $n_p$
bonds:
\begin{equation}
\mbox{e}^{-\lambda_{p} \frac{l_p}{v}}=P_{b_{n_p}b_{n_p-1}}\ldots
P_{b_{2}b_{1}} P_{b_{1}b_{n_p}}
\label{def.lyap.bis}
\end{equation}

We can thus write
\begin{equation*}
{\rm tr}\; {\mathsf Q}^{n}=\sum_{p\in
P_{n}}n_{p}\,\mbox{e}^{-\frac{\lambda_{p}}{v} r
l_{p}}\,\mbox{e}^{-\frac{s}{v}  r l_{p}}
\end{equation*}
where we have explicitly considered the degeneracy $n_{p}$ of the orbit due
to the number of points (vertices) from where the orbit can start.
Accordingly, we get
\begin{equation*}
\frac{{\rm tr}\; {\mathsf Q}^{n}}{n}=\sum_{p\in
P_{n}}\frac{1}{r}\,\mbox{e}^{-\frac{\lambda_{p}}{v} r
l_{p}}\,\mbox{e}^{-\frac{s}{v}rl_{p}}
\end{equation*}
The sum over $n$ of $\frac{{\rm tr}{\mathsf Q}^{n}}{n}$ is equivalent to
the sum over all
the periodic orbits and their repetitions, i.e.,
\begin{equation*}
\sum_{n=1}^{\infty}\frac{{\rm tr}\; {\mathsf Q}^{n}}{n} =
\sum_{p} \sum_{r=1}^{\infty}\frac{1}{r}\,\mbox{e}^{-(\lambda _{p} +
s) \frac{l_{p}}{v} r}
\end{equation*}
We recognize in the right-hand side the expansion of the logarithm,
whereupon
\begin{equation*}
\sum_{r=1}^{\infty} \frac{1}{r}\,\mbox{e}^{-(\lambda _{p} +
s) \frac{l_{p}}{v} r}=-\ln
\left[1-\mbox{e}^{-(\lambda _{p} +
s) \frac{l_{p}}{v}}\right]
\end{equation*}
We have thus
\begin{equation*}
\det ({\mathsf I}-{\mathsf Q})=\exp {\sum_{p}\ln \left[1-\mbox{e}^{-(\lambda _{p} + 
s) \frac{l_{p}}{v}}\right]}
\end{equation*}
or 
\begin
{equation}
\det ({\mathsf I}-{\mathsf Q})=\Pi_{p}\left[1-\mbox{e}^{-(\lambda _{p} +
s) \frac{l_{p}}{v}}\right]
\equiv Z(s)
\label{selbergsmale.graph}
\end{equation}
which is the Selberg-Smale zeta function for the classical dynamics on graphs.

\subsection{The Pollicott-Ruelle resonances}

The zeros of the zeta function (\ref
{selbergsmale.graph}) are the so-called Pollicott-Ruelle resonances.
The results here above show that the spectrum of the Pollicott-Ruelle
resonances controls the
asymptotic time evolution and the relaxation properties of the dynamics on
the graphs.
In general, the zeros $s_{j}$ are located in the half-plane
${\mathrm Re}s_{j}\leq 0$ because the density does not grow exponentially
in time.

The spectrum of the zeros of the Selberg-Smale zeta function allows us to
understand the main
features of the classical Liouvillian time evolution of a system.  Let us
compare the classical
zeta function  (\ref{selbergsmale.graph}) for graphs with similar classical
zeta functions
previously derived for deterministic dynamical systems
\cite{Cvitanovic,Gasp98}.  For Hamiltonian
systems with two degrees of freedom, the classical zeta function is given
by two products: (1) the
product over the periodic orbits as in the case (\ref{selbergsmale.graph})
of graphs; (2) an extra
product over an integer $m=1,2,3,...$ associated with the unstable
direction transverse to the
direction of the orbit.  This integer appears as an exponent of the factor
associated with each periodic orbit \cite{Gasp98}.  As a consequence of
this extra
product, some zeros of the zeta function are always degenerate for a reason
which is intrinsic to
the Hamiltonian dynamics of a system with two or more degrees of freedom.
Accordingly,
Jordan-block structures are possible in typical Hamiltonian systems.

In contrast, no such degeneracy of dynamical origin appears in classical
graphs because no integer
exponent affects the periodic-orbit factors in Eq.
(\ref{selbergsmale.graph}).  In general, this
property does not exclude the possibility of degenerate zeros which may
appear for reasons
of geometrical symmetry of a graph or for a particular choice of the
parameter values defining a
graph.  However, such degeneracies are not expected for typical values of
the parameters which are
the transition probabilities $P_{bb^{\prime}}$ and the lengths $l_b$.
Examples will be given below
that illustrate this point.  According to this observation, Jordan-block
structures should not be
expected in typical graphs.

Different behaviors are expected depending on whether the graph is finite
or infinite.

\subsubsection{Finite graphs}

Finite graphs are composed of a finite number of finite bonds.
In this case, the matrix ${\mathsf Q}(s)$ is finite of size $2B\times 2B$
with exponentials
$\exp(-sl_b/v)$ in each element.  The characteristic determinant and
therefore the zeta function
(\ref{selbergsmale.graph}) is thus given by a finite sum of terms with
exponential functions of
$s$.  As a consequence, the Selberg-Smale zeta function is an entire
function of exponential
type in the complex variable $s$:
\begin{equation*}
\vert Z(s)\vert \ \leq \ K \ \exp\left(\frac{L_{\rm tot}}{v}\vert s\vert\right)
\end{equation*}
where $K$ is a positive constant and $L_{\rm tot}=\sum_{b=1}^{2B}l_b$ is
the total length of the
directed graph (which is finite by assumption). Hence, the zeta function is
analytic and has neither pole
nor other singularities.  In general, such a zeta function only has
infinitely many zeros
distributed in the complex plane $s$.

The finite graphs form closed systems in which the particle always remains
at finite distance
without escaping to infinity. For closed systems, we should expect that
there exist equilibrium
states defined by some invariant measures.  Such equilibrium states are
reached after all the
transient behaviors have disappeared in the limit $t\to +\infty$.

According to the spectral decomposition (\ref{spectral}), the equilibrium
states should
thus correspond to vanishing relaxation rates $s_j=0$.  Whether the
equilibrium state is unique
or not is an important question.  In the affirmative, the system is ergodic
otherwise it is
nonergodic.  Because of the definition (\ref{Eq.Q}), we have that
$Q_{bb^{\prime}}(0)=P_{bb^{\prime}}$ so that the value $s=0$ is a root of
the characteristic
determinant (\ref{detI-Q}) if the matrix
${\mathsf P}$ of the transition probabilities admits the unit value as
eigenvalue.
Because of the condition (\ref{norm}), we know that the unit value is
always an eigenvalue of
$\mathsf P$.  The question is whether this eigenvalue is simple or not.  If
it is simple, the
equilibrium state is unique and the system ergodic otherwise it is multiple
and the system
nonergodic.

In order to answer the question of ergodicity, let us introduce the
following definition:
\begin{equation*}
{\mathsf P}\  \mbox{is {\it irreducible} iff}\quad  \forall b,b^{\prime}\;
, \quad \exists n\;
:\qquad ({\mathsf P}^n)_{bb^{\prime}} > 0
\end{equation*}
Then, we have the result that:

{\it The classical dynamics on a finite graph is ergodic if the matrix of
the transition
probabilities is irreducible.}

Indeed, if the transition matrix is irreducible all the bonds are
interconnected so that there
always exist $n$ transitions that will bring the particle from any bond
$b^{\prime}$ to any other
bond $b$.  It means that the graph is made of one piece, i.e., the dynamics
on the graph is
said to be {\it transitive}.  The irreducibility of the transition matrix
implies the unicity of
the equilibrium state because of the Frobenius-Perron theorem \cite{walters}:

{\it If a matrix has non-negative elements and is irreducible, there is a
non-negative and
simple eigenvalue which is greater than or equal to the absolute values of
all the other
eigenvalues.  The corresponding eigenvector and its adjoint have strictly
positive components.}

We notice that the transition matrix $\mathsf P$ is non-negative and that
no eigenvalue is
greater than one because all the matrix elements obey $0\leq
P_{bb^{\prime}} \leq 1$ and,
moreover, Eq. (\ref{norm}) holds. On the other hand, we know that the unit
value is an eigenvalue
also because of (\ref{norm}). Therefore, if the transition matrix is
assumed to be irreducible,
the eigenvalue $1$ is simple. According to Eq. (\ref{right.eigen}), the
equilibrium state of
relaxation rate $s_0=0$ is given by the unique positive eigenvector
$\pmb{\chi}_0={\mathsf
P}\cdot\pmb{\chi}_0={\mathsf Q}(0)\cdot\pmb{\chi}_0$ corresponding to the
simple eigenvalue $1$ as
\begin{equation}
\Psi_0[b,x] = \chi_0[b] \ , \qquad \mbox{for} \quad 0 < x<l_b
\end{equation}
The corresponding adjoint eigenvector of $\mathsf P$ is
$\tilde{\chi}_0[b]=1$, $\forall b$.
The positive component $\chi_0[b]$ of the right eigenvector gives the
probability to find the
particle in the bond $b$ at equilibrium.  These components obey the
probability normalization
$\sum_b\chi_0[b]=1$.  This equilibrium state defines an invariant
probability measure in the
space of trajectories:
\begin{equation}
\mu(b_{n-1}\cdots b_1b_0) = P_{b_{n-1}b_{n-2}} \cdots P_{b_2b_1}P_{b_1b_0}
\chi_0[b_0]
\label{inv.meas}
\end{equation}

Examples of nonergodic graphs are disconnected graphs.

The classical dynamics on a closed graph is said to be {\it mixing} if
there is no pure
oscillation in the asymptotic time evolution, i.e., if there is no
resonance with ${\rm Re}\;
s_j=0$ except the simple resonance $s_0=0$.  According to Eq.
(\ref{right.eigenstate}) we have for a mixing graph that
\begin{equation}
\lim_{t\to +\infty} \; \langle A\rangle_t = \sum_b \chi_0[b] \;
\frac{1}{l_b} \int_0^{l_b} A[b,x]
\; dx
\end{equation}

An example of a graph which is ergodic but
non-mixing is a single bond of length $g$ between two vertices.  Its zeta
function is
$Z(s)=1-\exp(-2sg/v)$ so that its resonances are
\begin{equation*}
s_j = i \; v \; \frac{\pi }{g}\;  j \, \qquad \mbox{with} \quad j\in{\mathbb Z}
\end{equation*}
Except $s_0=0$, all the other resonances are pure imaginary so that the
dynamics is oscillatory
as expected.

\subsubsection{Infinite graphs of scattering type}

\label{subsubsec.infinite}

Graphs of scattering type can be constructed by attaching semi-infinite
leads $c$ to a finite
graph. These semi-infinite leads are bonds of infinite length.
As soon as the particle exits the finite part of the graph by one of these
leads it escape in
free motion toward infinity, which is expressed by the vanishing of the
following probabilities
between the semi-infinite leads $c$ and every bond $b$ of the finite part
of the graph:
\begin{equation*}
P_{bc}=0\qquad \forall b
\end{equation*}
and
\begin{equation*}
P_{\hat{c}b}=0\qquad \forall b
\end{equation*}
Therefore, $Q_{bc}=Q_{\hat{c}b}=0$, $\forall b$, and
\begin{equation*}
\det ({\mathsf I}-{\mathsf Q})=\det ({\mathsf I}-\tilde{{\mathsf Q}})
\end{equation*}
where $\tilde{{\mathsf Q}}$ is the matrix for the finite part of the graph
without the
scattering leads.  Since the leads cause the particle to escape to
infinity, the probability
for the particle to stay inside the graph is expected to decay.  Therefore,
the zeros
of the Selberg-Smale zeta function are located in the  half-plane ${\mathrm
Re}\,s_{j}<0$
and there is a gap empty of resonances below the axis
${\mathrm Re}\,s=0$: ${\mathrm Re}\, s_j \leq s_0 < 0$.  The resonance
$s_0$ with the largest (or
smallest in absolute value) real part is real because the classical zeta
function is real.  This
leading resonance determines the exponential decay after long times which
we call the classical
escape rate $\gamma_{\rm cl}=-s_0$ (or the inverse of the classical
lifetime of a particle
initially trapped in the scattering region $\tau_{\rm cl}=1/\gamma_{\rm
cl}$). The trajectories
which remain trapped form what we shall call a repeller because it is the
analogue of the
repeller in deterministic dynamical systems with escape \cite{Gasp98}.

An invariant measure can be defined on this repeller by applying the
Frobenius-Perron operator to
the non-negative matrix $\tilde{\mathsf Q}(s_0)$ evaluated at the leading
resonance.  This matrix
has a leading eigenvalue equal to one and the corresponding left and right
eigenvectors are
positive.  A matrix of transition probabilities on the repeller can be
defined by
\begin{equation}
\Pi_{bb^{\prime}} = \tilde\chi_0[b] \; \tilde{Q}_{bb^{\prime}} \;
\frac{1}{\tilde\chi_0[b^{\prime}]}
\end{equation}
which leaves invariant the probabilities
\begin{equation}
\pi[b] = \frac{\chi_0[b] \;
\tilde\chi_0[b]}{\pmb{\chi}_0\cdot\tilde{\pmb{\chi}}_0}
\end{equation}
of finding the particle of each bond $b$ in its motion on the repeller.
These probabilities obey
\begin{equation}
\sum_b \Pi_{bb^{\prime}} = 1 \ , \qquad \mbox{and} \qquad \sum_{b^{\prime}}
\Pi_{bb^{\prime}} \; \pi[b^{\prime}] = \pi[b]
\end{equation}
and the invariant measure on the repeller is defined as
\begin{equation*}
\mu(b_{n-1}\cdots b_1b_0) = \Pi_{b_{n-1}b_{n-2}} \cdots
\Pi_{b_2b_1}\Pi_{b_1b_0} \pi[b_0]
\end{equation*}

As an example, consider the graph
formed by one bond of length $g$ that joins two vertices and
two scattering leads attached to one of these vertices. The repeller consists
here only of one unstable periodic orbit. Thus we look for the complex
solutions of
\begin{equation*}
1-\mbox{e}^{-(\lambda _{p} + s) \frac{l_{p}}{v}}=0
\end{equation*}
that is
\begin{equation*}
\lambda _{p}+s=-i \; v\; \frac{2\pi}{l_p} \; j \ , \qquad \mbox{with} \qquad
j\in{\mathbb Z}
\end{equation*}
where $l_{p}=2g$ and $(l_{p}\lambda _{p}/v)=-\ln (1/9)$ which follows from
Eqs. (\ref{Ts}) and (\ref{def.lyap}). Accordingly, we get
\begin{equation*}
s_{j}=-v\; \frac{\ln 9}{2g} + i \; v \; \frac{\pi}{g} \; j \ , \qquad
\mbox{with} \qquad
j\in{\mathbb Z}
\end{equation*}
Therefore, all the resonances have the lifetime $\tau _{\rm
cl}=\frac{2g}{v\ln 9}=
\frac{g}{v\ln 3}$. This lifetime coincides with the quantum lifetime
obtained from the resonances of the same graph \cite{Felipe4}.
Another system having this peculiarity is the two-disk scatterer
\cite{Gasp98}. This
property is due to the fact that there is only one periodic orbit. In the
presence of chaos and thus infinitely many periodic orbits, the quantum
lifetimes
are longer than the classical ones \cite{Felipe4,GaspRice}.

\section{The chaotic properties}

\label{sec.chaotic}

\subsection{Correspondence with deterministic chaotic maps}

The previous results show that the classical dynamics on a graph is random.
It turns out that
this dynamical randomness is not higher than the dynamical randomness of a
deterministic chaotic
system.

In order to demonstrate this result, we shall establish the correspondence
between the
random classical dynamics on a graph and a suspended flow on a
deterministic one-dimensional map
of a real interval.  As aforementioned, the trajectories of the random
dynamics on a graph are in
one-to-one correspondence with bi-infinite sequences giving the directed
bonds successively
visited by the particle, $\cdots b_{-2}b_{-1}b_{0}b_{1}b_{2}\cdots $, which
is composed of
integers $1\leq b_n\leq 2B$.  For simplicity, we shall only consider the
future time evolution
given by the infinite sequence $b_{0}b_{1}b_{2}\cdots $.  With each
infinite sequence, we can
associate a real number in the interval $0\leq y \leq 2B$ thanks to the
formula for the $2B$-adic
expansion
\begin{equation}
y = \sum_{n=0}^{\infty} \frac{(b_n-1)}{(2B)^n}
\end{equation}
Accordingly, the directed bond $b'$ is assigned to the subinterval
$b'-1<y<b'$.  Each of these
subintervals is subdivided into $2B$ smaller subintervals
\begin{multline}
Y_{b-1,b'}<y<Y_{b,b'} \ , \\ \mbox{with} \quad Y_{b,b'}=Y_{b-1,b'}+P_{bb'}
\ , \quad
Y_{0,b'}=b'-1\ , \quad \mbox{and} \quad Y_{2B,b'}=b'
\end{multline}
The one-dimensional map is then defined on each of these small subintervals
by the following
piece-wise linear function
\begin{equation}
y_{n+1} = \phi(y_n) \equiv \frac{1}{P_{bb'}} \left( y_n- Y_{b-1,b'}\right)
+ b-1 \ , \qquad
\mbox{for} \quad Y_{b-1,b'}<y_n<Y_{b,b'} \label{map}
\end{equation}
Since the transition probabilities are smaller than one, $0\leq P_{bb'}\leq
1$, the slope of
the map is greater than one: $1\leq\frac{d\phi}{dy}$.  As a consequence,
the map (\ref{map}) is
in general expanding and sustains chaotic behavior.

The suspended flow is defined over this one-dimensional map with the
following return-time
function giving the successive times $t_n$ of return in the surface of section:
\begin{equation}
t_{n+1} = t_n + T(y_n)
\end{equation}
with
\begin{equation}
T(y_n) \equiv \frac{l_{b'}}{v} \ , \qquad \mbox{for} \quad
Y_{b-1,b'}<y_n<Y_{b,b'}
\end{equation}

For finite and closed graphs, the invariant measure of the one-dimensional
map (\ref{map}) is
equal to
\begin{equation}
\rho_{\rm eq}(y) = p_{b'} \ , \qquad \mbox{for} \quad b'-1<y<b'
\end{equation}

For infinite graphs of scattering type, the function (\ref{map}) maps the
subintervals associated
with the semi-infinite leads outside the interval $0\leq y \leq 2B$,
generating an escape process.
For such open graphs, the one-dimensional map selects a set of initial
conditions of
trajectories which are trapped forever in the interval $0\leq y \leq 2B$.
This set of zero
Lebesgue measure is composed of unstable trajectories and is called the
repeller.  Typically,
this repeller is a fractal set.

We notice that, for closed graphs, an isomorphism can even be established
between the dynamics in
the space of bi-infinite sequences and a two-dimensional area-preserving
map according to a
construction explained elsewhere \cite{Gasp98,GaspWang}.

\subsection{Characterization of the chaotic properties}

The chaotic properties can be characterized by quantities such as the
topological
entropy, the Kolmogorov-Sinai entropy, the mean Lyapunov exponent, or the
fractal dimensions in
the case of open systems.  All these quantities can be derived from the
so-called ``topological
pressure" $P(\beta)$ \cite{GaspDorfman}.  This pressure can be defined per
unit time or
equivalently per unit length since the particle moves with constant
velocity $v$ on the graph.

The topological pressure can be defined in analogy with the definition
for time-continuous systems. For this goal, we notice that time is related
to length by
$v=l/t$ and that the role of the stretching factors is played by the
inverses of the transition
probabilities in the context of graphs.  Accordingly, the topological
pressure per unit time is
defined by
\begin{equation}
P(\beta ) \equiv \lim_{L\rightarrow \infty }\frac{v}{L}\ln \sum_{\substack{
b_{0}\cdots b_{n-2}  \\ L<l_{b_0}+\cdots + l_{b_{n-2}}<L+\Delta L}}\left(
P_{b_{n-1}b_{n-2}}\cdots
P_{b_{1}b_{0}}\right) ^{\beta }
\label{pressure}
\end{equation}
where the sum is restricted to all the trajectories that remain in the graph
and do not escape (i.e., on the repeller) and that have a length that
satisfies $
L<l_{b_0}+\cdots + l_{b_{n-2}}<L+\Delta L$ (cf. Refs.
\cite{Gasp98,GaspDorfman}). The dependence on
$\Delta L$ disappears in the limit $L\to\infty$.

The equation (\ref{pressure}) can be expressed by the condition that the
pressure is given by
requiring that the following sum is approximately equal to one in the limit
$n\to\infty$
\begin{equation}
1 \sim_{n\to\infty} \sum_{b_{0}\cdots b_{n-2}} \left( P_{b_{n-1}b_{n-2}}\cdots
P_{b_{1}b_{0}}\right)^{\beta} \; \mbox{e}^{-\frac{1}{v}P(\beta)
\left(l_{b_{n-2}} +\cdots +
l_{b_0}\right)}
\label{pressure.bis}
\end{equation}
which is equivalent to requiring that the matrix ${\mathsf Q}(s;\beta)$
composed of the elements
\begin{equation}
Q_{bb'}(s;\beta) \equiv (P_{bb'})^{\beta} \; \mbox{e}^{-s \frac{l_{b'}}{v}}
\label{pressure.Q}
\end{equation}
with $s=P(\beta)$ has the eigenvalue $1$ as its largest eigenvalue.
As a consequence, the topological pressure can be obtained as the leading
zero of the following
zeta function
\begin{equation}
Z(s;\beta) = \det\left[ {\mathsf I}-{\mathsf Q}(s;\beta)\right]
\label{pressure.zeta}
\end{equation}
or, equivalently, as the leading pole $s=P(\beta )$ of the Ruelle zeta function
\begin{equation}
\zeta_{\beta}(s) \equiv \frac{1}{Z(s;\beta)} = \prod_p
\frac{1}{1-\mbox{e}^{-(\beta
\lambda_{p}+s)\frac{l_p}{v} } }.  \label{eq.zeta-beta}
\end{equation}

The different characteristic quantities are then determined in terms of the
topological
pressure function as follows \cite{Gasp98}:

\begin{itemize}
\item  The escape rate is given by $\gamma_{\rm cl} =-P(1)$;

\item  The mean Lyapunov exponent by $\lambda =-P^{\prime }(1)$;

\item  The Kolmogorov-Sinai entropy is determined by $h_{\rm KS}=\lambda
-\gamma_{\rm cl} = P(1)-P^{\prime }(1)$;

\item  The topological entropy by $h_{\rm top} = P(0)$;

\item  The Hausdorff partial dimension of the repeller of the corresponding
one-dimensional map
(\ref{map}) is the zero of $P(\beta)$, i.e., $P(d_{\rm H})=0$.

\end{itemize}

The mean Lyapunov exponent, the escape rate, and the entropies are defined
per unit time.  The
mean Lyapunov exponent characterizes the dynamical instability due to the
branching of the
trajectories on the graph.  On the other hand, the KS entropy characterizes
the global dynamical
randomness.  Both would be equal if the graph was closed and the escape
rate vanished.  We shall
say that {\em the dynamics on a graph is chaotic if its KS entropy is
positive, $h_{\rm KS}>0$}.
We emphasize that a dynamics with a positive Lyapunov exponent is not
necessarily chaotic.  A
counterexample to this supposition is given by the open graph at the end of
the previous
subsection.  The repeller of this graph is composed of a single periodic
orbit and its Lyapunov
exponent is equal to the escape rate:
$\lambda=\gamma_{\rm cl}=(v\ln 3)/g$.  Accordingly, its KS entropy vanishes
in agreement with the
periodicity of this dynamics.

We notice that the escape rate is related to the leading Pollicott-Ruelle
resonance by
$\gamma_{\rm cl}=-s_0$.  Indeed, when $\beta=1$ the zeta function
(\ref{pressure.zeta}) reduces
to the previous one given by Eq. (\ref{selbergsmale.graph}) which has the
Pollicott-Ruelle
resonances as its zeros.

Moreover, we have the following properties:

\begin{itemize}

\item[1.]{\em The topological entropy is independent of the transition
probabilities $P_{bb'}$
of the graph.}

\item[2.]{\em The Hausdorff dimension is independent of the lengths $l_b$
of the graph.}

\end{itemize}

The first property is deduced from Eq. (\ref{eq.zeta-beta}) when we set
$\beta=0$ to calculate the
topological entropy.  In this case, we observe that the Lyapunov exponents
disappear from the
zeta function (\ref{eq.zeta-beta}) which thus depends only on the
lengths of the periodic orbits of the graph.  As a consequence, the
topological entropy, which is
given by the leading pole of the Ruelle zeta function (\ref{eq.zeta-beta})
with $\beta=0$, depends
only on the lengths of the bonds.

The second property can be inferred from Eq. (\ref{pressure.bis}) or,
similarly, from the
characteristic determinant (\ref{pressure.zeta}) for the matrix
(\ref{pressure.Q}).  Indeed,
since the Hausdorff dimension is the zero of the pressure function the
lengths now disappear when
we set $s=P(\beta)=0$ in either Eq. (\ref{pressure.bis}) or Eq.
(\ref{pressure.Q}).  Accordingly,
the Hausdorff dimension depends only on the transition probabilities.

\begin{figure}[t]
\centering
\includegraphics[width=8cm]{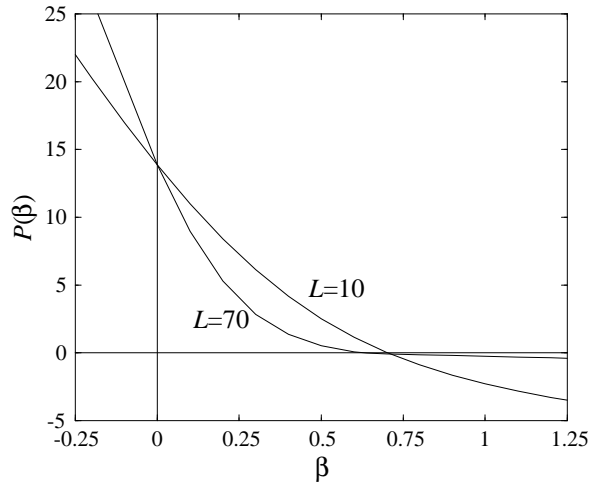}
\caption{The topological pressure for a fully connected pentagon
with $L=10$ and $L=70$ leads attached to each vertex. The velocity is $v=1$.}
\label{fig.prespenta}
\end{figure}

\begin{figure}[t]
\centering
\includegraphics[width=10cm]{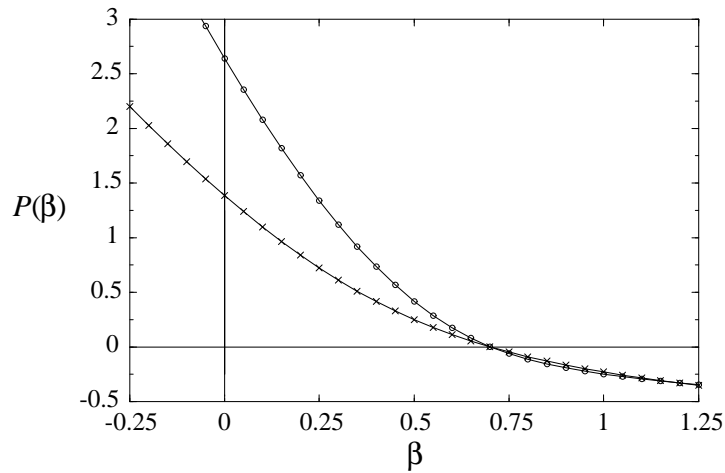}
\caption{The topological pressure for a fully connected pentagon
with $L=10$. The curve with the crosses is obtained when all bonds as
unit length and the curve with the circles for a set of incommensurate lengths.
The velocity is $v=1$.}
\label{fig.pentapres}
\end{figure}


\section{Scattering on open graphs}
\label{sec.scatt}

We shall consider some examples that illustrate the previous concepts.
Consider the fully connected pentagon with $L$ scattering leads attached
to each vertex. Since the topological entropy $h_{\rm top}$ is independent
of the Lyapunov exponents it is independent of the number of
scattering leads attached to each vertex. This is observed in
Fig. \ref{fig.prespenta} where
we depict the topological pressure for the fully connected pentagon.

Moreover, we observe that the escape rate $\gamma_{\rm cl}=-P(1)$
for the pentagon with $L=70$ is smaller
than the escape rate for the pentagon with $L=10$. This behavior has a simple
interpretation.  Since we use
$P_{bb'}=\left|\frac{2}{\nu_{bb'}}-\delta_{\hat{b'}b'}\right|^2$,
the transmission probability from bond to bond decreases
and the reflection probability increases as the valence
of the vertex $\nu_{bb'}$ increases. Therefore, as the number of leads
increases,
a particle on the pentagon has a smaller probability to escape
and a larger probability to be reflected back to the same bond. Accordingly,
the escape rate diminishes.

The examples of Fig. \ref{fig.pentapres} shows that, indeed, the
Hausdorff dimension is independent of the bond lengths.

As we see in these examples, the dynamics on typical graphs is
characterized by
a positive KS entropy $h_{\rm KS}>0$. In this sense, the classical dynamics
on typical graphs is chaotic.


\section{Diffusion on infinite periodic graphs}

\label{sec.diff}

\subsection{The hydrodynamic modes of diffusion}

If the evolution of the density in an infinite periodic graph corresponds to
a diffusion process, then the phenomenological diffusion equation should be
satisfied in some limit. For instance, if the periodic graph forms a chain
extending from $x=-\infty$ to $x=\infty$ then on a large scale (much larger
than the period of the system) the density profile should evolve according to
the diffusion equation

\begin{equation}
\frac{\partial\rho }{\partial t}=D \frac{\partial^2 \rho}{\partial x^2}
\label{diffusion.eq}
\end{equation}
Let us notice that $x$ is a one-dimensional coordinate of position along the
graph which is {\em a priori} different from the position along each bond.

This equation admits solutions of the form
\begin{equation*}
\rho_k=\exp\left[s(k)\; t\right]\exp (ikx)
\end{equation*}
with the dispersion relation
\begin{equation}
s(k)=-Dk^{2}  \label{diff.disp.rel}
\end{equation}
that relates the eigenvalue $s_k$ to the wave number $k$.  These
solutions are called the hydrodynamic modes of diffusion.
The inverse of the wave number gives the wavelength $\ell =2\pi/k$ of the
spatial
inhomogeneities of concentration of particles.

For a system such as a graph, we expect deviations with respect to the
diffusion
equation which only gives the large-scale behavior of the probability density
and not the behavior on the scale of the bonds.
Moreover, we may also expect the existence of other kinetic modes of faster
relaxation than the leading diffusive hydrodynamic mode.
In order to obtain a full description of the relaxation,
we have to compute the eigenvalues of the evolution operator
for an infinite periodic graph.  One of
these eigenvalues will have the dependence of Eq. (\ref{diff.disp.rel})
for small $k$ which allows
us to obtain from it the diffusion coefficient of the chain. We shall start by
considering periodic graphs in a $d$-dimensional space to show the
generality of the
method but then we will specialize to periodic graphs that form
one-dimensional chains.

\subsection{Fourier decomposition of the Frobenius-Perron operator}

In spatially extended systems which form a periodic lattice, spatial Fourier
transforms are needed in order to reduce the dynamics to an elementary cell
of the lattice \cite{GaspEigenDiff}. In this reduction a wave number $\bf
k$ is introduced
for each hydrodynamic mode.  The wave number characterizes a spatial
quasiperiodicity of the probability density with respect to the lattice
periodicity.  Indeed, the wavelength of the mode does not need to be
commensurate with the size of a unit cell of the lattice.  Each Fourier
component
of the density evolves independently with an evolution operator that
depends on $\bf k$.
Accordingly, the Pollicott-Ruelle resonances will also depend on
$\bf k$. We shall implement this reduction starting from the
master equation (\ref{Ev.t}) of the fully periodic graph and construct
from it the evolution operator for the unit cell. A few new definitions
are needed before proceeding with this construction.

The periodic graph is obtained by successive repetitions of a unit cell.
Such graphs form a Bravais lattice $\mathcal{L}$. A lattice
vector ${\bf a}_{\bf m}$ is centered in each cell of the Bravais
lattice with ${\bf m}=(m_{1},m_{2},...,m_{d})\in {\mathbb Z}^{d}$.
The lattice vectors are given by linear combinations of the basic vectors
of the lattice
\begin{equation*}
{\bf a}_{\bf m}=m_{1}{\bf a}_{100\cdots 00}+m_{2}{\bf a}_{010\cdots 00}+\cdots
+ m_{d}{\bf a}_{000\cdots 01}\in
\mathcal{L}
\end{equation*}
(for a one-dimensional chain we have $d=1$ and $a_{m}=m\in{\mathbb Z}$). We
shall split the
coordinate $[b,x ]$ which refers to an arbitrary bond in the infinite
graph into the new coordinates $([\mathrm{b},x ],{\bf a})$ where the
first pair refers to the equivalent position in the elementary unit cell to
which the dynamics is reduced. That is, the bond $\mathrm{b}$ is associated
with
$b$ and $x$ is the position in that bond. The third term represents a vector
in the Bravais lattice that gives the true position of the bond $b$ with
respect to the position of the original unit cell, that is $b$ is obtained
by translating the bond \textrm{b} to the cell in the Bravais lattice
identified by the vector ${\bf a}$. We may introduce the notation
$b=T_{\bf a}(\mathrm{b})$ where the translation operators $T_{\bf a}$
assign to a bond
$\mathrm{b}$ the corresponding bond in the unit cell characterized by the
vector ${\bf a}$.

Accordingly, the density in the graph is represented by a new function $
\footnote{
We keep calling $\rho $ this new density.}$ $\rho$ related to the old one
by $\rho ([{\mathrm{b}},x],{\bf a},t)=\rho([b,x ],t)$.

We define a projection operator by
\begin{equation}
\hat{E}_{\bf k}=\sum_{{\bf a}\in \mathcal{L}}\exp (-i{\bf k}
\cdot {\bf a})\hat{S}^{\bf a},  \label{projector}
\end{equation}
in terms of the spatial translation operators
\begin{equation*}
\hat{S}^{\bf a}f([{\mathrm{b}},x ],{\bf a}^{\prime},t)=
f([{\mathrm{b}},x ],{\bf a}^{\prime }+{\bf a},t)\quad
\text{for}\ {\bf a},{\bf a}^{\prime} \in \mathcal{L}.
\end{equation*}
The projection operator (\ref{projector}) involves the so-called wave number
${\bf k}$. This later is defined on the Brillouin zone $\mathcal{B}$ of the
reciprocal lattice $\widetilde{\mathcal{L}}$. The volume of the Brillouin
zone is
\begin{equation*}
|{\mathcal{B}}|=\int_{{\mathcal{B}}}d{\bf k}=\frac{(2\pi )^{d}}{\left| \det (
{\bf a}_{100\cdots 00},{\bf a}_{010\cdots 00},...,{\bf a}_{000\cdots
01})\right| }.
\end{equation*}
The operators (\ref{projector}) are projection operators since
\begin{equation*}
\hat{E}_{\bf k}\hat{E}_{{\bf k}^{\prime}}=|\mathcal{B}
|\; \delta ({\bf k}-{\bf k}^{\prime })\hat{E}_{\bf k}
\end{equation*}
which is a consequence of the relation
\begin{equation*}
\frac{1}{|\mathcal{B}|}\sum_{{\bf a}\in\mathcal{L}}\exp (i{\bf k}
\cdot {\bf a})=\sum_{{\bf k}^{\prime}\in \widetilde{\mathcal{L}}
}\; \delta ({\bf k}-{\bf k}^{\prime}).
\end{equation*}
The identity operator is recovered by integrating the projection operator
over the wave number
\begin{equation*}
\hat{I}=\frac{1}{|\mathcal{B}|}\int d{\bf k}\; \hat{E}_{\bf k}.
\end{equation*}
If $\rho$ is the density defined on the infinite phase space, the function
$\hat{E}_{\bf k}\rho $ is quasiperiodic on the lattice
\begin{multline*}
\hat{E}_{\bf k}\rho ([{\mathrm{b}},x ],{\bf a},t)=
\sum_{{\bf a}^{\prime }\in {\mathcal{L}}}\exp (-i{\bf k}\cdot
{\bf a}^{\prime})\rho ([{\mathrm{b}},x ],{\bf a}+{\bf a}^{\prime },t) \\
= \exp (i{\bf k}\cdot {\bf a})\sum_{{\bf a}^{^{\prime \prime }}\in
{\mathcal{L}}}\exp (-i{\bf k}\cdot {\bf a}^{^{\prime \prime }})\rho ([
{\mathrm{b}},x ],{\bf a}^{^{\prime \prime }},t) \\ =\exp (i{\bf k}\cdot
{\bf a})\hat{E}_{\bf k}\rho ([{\mathrm{b}},x ],{\bf 0},t)
=\exp (i{\bf k}\cdot {\bf a})\rho_{\bf k}([{\mathrm{b}},x ],t).
\end{multline*}
We have therefore a decomposition of the density over the infinite phase
space into components defined in the reduced phase space and which depends
continuously on the wave number ${\bf k}$.

Consider the master equation (\ref{Ev.t}) of the full periodic graph.
Applying at both sides the operator $\hat{E}_{\bf k}$ we get from
the quasiperiodicity of $\hat{E}_{\bf k}\rho$ [see the previous
equation]
\begin{equation}
\exp (i{\bf k}\cdot {\bf a}_{{\bf m}})\rho_{\bf k}([{\mathrm{b}}
,x ],t)=\sum_{b^{\prime }}P_{bb^{\prime }}\exp (i{\bf k}\cdot
{\bf a}_{{\bf m}^{\prime }})
\rho_{\bf k}\left([{\mathrm{b}}^{\prime },x ],t-\frac{x+
l_{{\mathrm{b}}^{\prime }}-x^{\prime}}{v}\right) .
\label{M.eq.proj}
\end{equation}
where $b=T_{{\bf a}_{\bf m}}({\mathrm{b}})$ and $b^{\prime }=T_{
{\bf a}_{{\bf m}^{\prime }}}({\mathrm{b}}^{\prime })$ and also $l_{b}=l_{
{\mathrm{b}}}$. Now, the translational symmetry implies that
\begin{equation*}
P_{bb^{\prime }}=P_{T_{{\bf a}_{\bf m}}({\mathrm{b}}),T_{{\bf a}_{
{\bf m}^{\prime }}}({\mathrm{b}}^{\prime })}=P_{T_{{\bf a}_{\bf
m-m'}}({\mathrm b}),{\mathrm{b}}^{\prime }}
\end{equation*}
Thus Eq. (\ref{M.eq.proj}) is
\begin{equation*}
\rho_{{\bf k}}([{\mathrm{b}},x ],t)=\sum_{{\mathrm{b}}^{\prime }}\sum_{
{\bf m}^{\prime }}P_{T_{{\bf a}_{\bf m-m'}}({\mathrm
b}),{\mathrm{b}}^{\prime }}
\; \exp (-i{\bf k}\cdot {\bf a}_{{{\bf m}}-{{\bf m'}}}) \
\rho _{{\bf k}}\left( [{\mathrm{b}}
^{\prime },x ],t-\frac{x+l_{{\mathrm{b}}^{\prime }}-x^{\prime } }{v}
\right)
\end{equation*}
The time evolution of the $\bf k$-component is therefore controlled by the
matrix of elements
\begin{equation}
{\mathrm{P}}_{\mathrm{bb}^{\prime }}({\bf k})\equiv \sum_{{\bf
m}}P_{T_{{\bf a}_{\bf m}}({\mathrm
b})  ,{\mathrm{b}}^{\prime }}\exp (-i{\bf k}\cdot
{\bf a}_{{\bf m}})  \label{pre.Trans.periodic}
\end{equation}
We observe that, for a graph, there is at most one term in the sum of Eq.
(\ref{pre.Trans.periodic}).  Indeed, the coefficient $P_{T_{{\bf a}_{\bf
m}}({\mathrm b})
,{\mathrm{b}}^{\prime }}$ does not vanish if and only if the bonds $T_{{\bf
a}_{\bf m}}({\mathrm
b})$ and ${\mathrm{b}}^{\prime }$ are connected with the same vertex of the
infinite graph.
Furthermore, there is one and only one translation $T_{{\bf a}_{\bf m}}$
for which $T_{{\bf a}_{\bf
m}}({\mathrm b})$ and ${\mathrm{b}}^{\prime }$ are connected with the same
vertex.  Accordingly,
there exists a unique lattice vector ${\bf a}({\rm b},{\rm b}^{\prime})$
such that
\begin{equation}
{\mathrm P}_{\mathrm{bb}^{\prime }}({\bf k}) = P_{T_{ {\bf a}({\rm b},{\rm
b}^{\prime}) }
({\rm b})  ,{\rm b}^{\prime } }
\exp \left[ -i{\bf k} \cdot {\bf a}({\rm b},{\rm b}^{\prime }) \right]
\label{Trans.periodic}
\end{equation}
Thanks to this matrix,
we have that each Fourier component of the density evolves with an equation
\begin{equation}
\rho_{{\bf k}}([{\mathrm{b}},x ],t)=\sum_{{\mathrm{b}}^{\prime }}{\mathrm{P}
}_{{\mathrm{bb}}^{\prime }}({\bf k})\; \rho _{{\bf k}}\left( [{\mathrm{b}}
^{\prime },x ],t-\frac{x+l_{{\mathrm{b}}^{\prime }}-x ^{\prime }}{v}
\right)  \label{M.Eq.periodic}
\end{equation}
Here the sum is carried out over all the directed bonds of the unit cell
and the matrix
${\mathsf P}({\bf k})$ of elements $\mathrm{P}_{\mathrm{bb}^{\prime }}({\bf
k})$
defined by Eq. (\ref{Trans.periodic}) is a square matrix with dimension
equal to the number of directed bonds in the unit cell.

\subsection{The eigenvalue problem and the diffusion coefficient}

To study the eigenvalues of the Frobenius-Perron operator $\hat{R}_{\bf
k}^{t}$ defined by Eq.
(\ref{M.Eq.periodic}) we proceed as in Subsec. \ref{subsec.master}.  We
introduce the following
definitions
\begin{equation*}
\rho_{{\bf k},s}\left[ {\mathrm b}\right] =\int_{0}^{\infty }\mbox{e}^{-st}\rho
_{\bf k}\left( \left[ {\mathrm b},0\right] ,t\right) dt
\end{equation*}
and
\begin{equation}
Q_{{\mathrm b}{\mathrm b}^{\prime }}(s,{\bf k})={\mathrm P}_{{\rm bb}
^{\prime }}({\bf k})\; \exp \left(-s\frac{l_{{\mathrm b}^{\prime }}}{v}\right)
\label{defQ}
\end{equation}
for the elements of the matrix ${\mathsf Q}(s,{\bf k})$. Then, taking the
Laplace transform of Eq. (\ref{M.Eq.periodic}) we get
\begin{equation*}
\rho _{{\bf k},s}\left[ {\mathrm b}\right] =\sum_{{\mathrm b}^{\prime }}
Q_{{\mathrm b}{\mathrm b}^{\prime }}(s,{\bf k})\; \rho _{{\bf k},s}\left[
{\mathrm b}^{\prime }\right]
\end{equation*}
which has a solution only if the following classical zeta function vanishes
\begin{equation}
Z(s;{\bf k}) \equiv \det [{\mathsf I}-{\mathsf Q}(s,{\bf k})]=\prod_p
\left[1-\mbox{e}^
{-(\lambda_p +s)\frac{l_p}{v}-i{\bf k}\cdot{\bf a}_p}\right]=0
\label{Q.periodic}
\end{equation}
where the product extends over the prime periodic orbits $p$ of
the unit cell of the graph and where ${\bf a}_p=\sum_{i=1}^{n_p}
{\bf a}({\rm b}_{i+1},{\rm b}_i)$
is the displacement on the lattice along the periodic orbit
$p={\mathrm b}_1{\mathrm b}_2\cdots{\mathrm b}_{n_p}$ of prime period
$n_p$.\footnote{Here, ${\bf a}({\rm b}_{i+1},{\rm b}_i)$ denotes the jumps
over the
lattice during the transition
between the bond ${\mathrm b}_i$  and the bond ${\mathrm b}_{i+1}$.
We have to consider that ${\bf
a}({\rm b}_{i+1},{\rm b}_i)=0$ for transitions between bonds in the same
unit cell.}
From Eq. (\ref{Q.periodic}), we obtain the functions
$s_{j}({\bf k})$ and the corresponding
eigenstates $\pmb{\chi}_{j,{\bf k}}$ which satisfy
\begin{equation*}
\pmb{\chi}_{j,{\bf k}}={\mathsf Q}\left[s_{j}({\bf k}),{\bf k}\right]\cdot
\pmb{\chi}_{j,{\bf k}}.
\end{equation*}
Eq. (\ref{Q.periodic}) shows that the Pollicott-Ruelle resonances
$s_j({\bf k})$ are the zeros of a new classical Selberg-Smale
zeta function defined for the spatially periodic graphs.
The eigenvalues and eigenstates of the Frobenius-Perron operator
$\hat{R}_{{\bf k}}^{t}$ defined by Eq. (\ref{M.Eq.periodic}) are
constructed in the same way as in Subsec. \ref{subsec.master}. If we denote by
$\Psi_{j,{\bf k}}[{\mathrm b},x]$ the eigenstates of the flow, the
results are
\begin{equation*}
\hat{R}_{{\bf k}}^{t}\Psi_{j,{\bf k}}[{\mathrm b},x ]=\mbox{e}^{s_{j}(
{\bf k})t}\Psi _{j,{\bf k}}[{\mathrm b},x ]
\end{equation*}
with
\begin{equation*}
\Psi_{j,{\bf k}}[{\mathrm b},x ]=\mbox{e}^{-s_{j}({\bf k})\frac{x }{v}
}\chi_{j,{\bf k}}[{\mathrm b}] \qquad \mbox{for}\quad  0<x <l_{{\mathrm b}}
\end{equation*}
where $s_{j}({\bf k})$ is a solution of Eq. (\ref{Q.periodic}).

For ${\bf k}=0$, Eq. (\ref{M.Eq.periodic}) represents the evolution of the
density in the unit cell with periodic boundary conditions. The periodic
boundary conditions transform two bonds in a loop. In this way, for ${\bf
k}=0$,
we are studying the evolution of a closed graph and we saw in Subsec.
\ref{subsec.master} that, for a closed graph, the value $s=0$ is a solution
of Eq. (\ref
{Q.periodic}) and the associated eigenstate corresponds to the invariant
measure or equilibrium probability which is given by the $2B$-vector ($B$
being the number of bonds in the unit cell)
\begin{equation}
\pmb{\chi}_{0}=\frac{1}{2B}(1,1,\ldots ,1).  \label{eigen.norm}
\end{equation}
We may thus expect that, for $\bf k$ small enough, there exists a zero $s_{0}(
{\bf k})$ of Eq. (\ref{Q.periodic}) and a corresponding eigenstate
\begin{equation}
\pmb{\chi}_{0,{\bf k}}={\mathsf Q}\left[s_{0}({\bf k}),{\bf k}\right] \cdot
\pmb{\chi}_{0,
{\bf k}}  \label{eigen.periodic}
\end{equation}
such that $s_{0}({\bf k})_{\overrightarrow{{\bf k}\rightarrow 0}}0.$
This particular resonance can be identified with the dispersion relation of
the hydrodynamic mode of diffusion since this latter is known to vanish at
${\bf k}=0$ as
\begin{equation*}
s_{0}({\bf k})=-\sum_{\alpha\beta} D_{\alpha \beta}k_{\alpha}k_{\beta} +
{\mathcal{O}}({\bf k}^4)
\end{equation*}
so that the diffusion matrix is obtained as
\begin{equation}
D_{\alpha \beta }=-\left. \frac{1}{2}\; \frac{\partial ^{2}s_{0}({\bf k})}{
\partial k_{\alpha }\partial k_{\beta }}\right| _{{\bf k
}=0}  \label{diff.coef}
\end{equation}

\subsection{Example of infinite graph with diffusion}

We illustrate the results of this section with a very simple example of a
one-dimensional chain. Consider the graph used in Subsubsec. \ref
{subsubsec.infinite}.  This graph of scattering type can be used as a unit cell
for a periodic graph. The right lead is connected with the left lead of an
equivalent graph and so on. This graph looks like an infinite comb. The unit
cell can be considered as composed by two bonds, say $b$ and $a$. The
bond $b$ connects the dead vertex $2$ with the vertex $1$ and the bond $a$
connects the vertex $1$ with the vertex $1$ of the next cell. Thus the
valence of the vertices are $v_{2}=1$ and $v_{1}=3$ respectively. The
transition probabilities $P_{bb^{\prime }}$ are given by
\begin{equation*}
P_{bb^{\prime }}=\left| T_{bb^{\prime }}\right| ^{2}
\end{equation*}
with, $P_{b\hat{b}}=1,P_{ab}=\frac{4}{9},P_{\hat{b}b}=\frac{1}{9}
,P_{a\hat{a}}=\frac{1}{9},P_{\hat{b}\hat{a}}=\frac{4}{9},P_{
\hat{a}a}=\frac{1}{9}$ and zero otherwise. We first construct the matrix
${\mathrm P}(k)$. We note from its definition in Eq. (\ref{Trans.periodic})
that ${\mathrm{P}}_{\mathrm{bb}^{\prime }}(k)=P_{{bb}^{\prime }}$ for
bonds that belong to the unit cell. The $k$-dependent factors come
from bonds that connect consecutive cells. These bonds are:

\begin{itemize}

\item The bond $a$ of the cell at the left-hand side characterized by
$a_{m}=-1$
is connected with the bonds $a$ and $\hat{b}$. This gives the contributions
${\mathrm P}_{{\rm aa}}(k)=\frac{4}{9}
\mbox{e}^{+ik}$ and ${\mathrm P}_{\hat{\rm b}{\rm a}}(k)=
\frac{4}{9}\mbox{e}^{+ik}$.

\item The bonds $\hat{a}$ and $b$ of the cell at the right-hand side
characterized by $a_{m}=+1$ are connected with the bond $\hat{a}$.
This gives the contributions ${\mathrm{P}}_{\hat{\rm a}
\hat{\rm a}}(k) =\frac{4}{9}\mbox{e}^{-ik}$ and ${\mathrm{P}}_{\hat{\rm
a}{\rm b}}(k)
=\frac{4}{9}\mbox{e}^{-ik}$.

\end{itemize}

\begin{figure}[t]
\centering
\includegraphics[width=9cm]{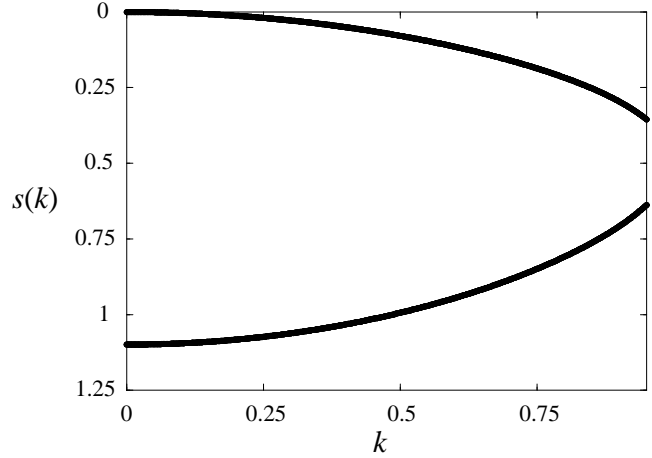}
\caption{The first two branches $s(k)$ of Pollicott-Ruelle resonances of
the infinite comb graph, as
obtained from Eq. (\ref{detQ.example}). The branch containing the origin
$s=0$ and $k=0$ is the
dispersion relation of the diffusive mode.  The other branch is associated
with a kinetic mode
of faster relaxation.}
\label{fig.ex.dif}
\end{figure}

Therefore, ${\mathsf P}(k)$ is the $4\times 4$ matrix with entries
\begin{equation*}
{\mathsf P}(k)=\left[
\begin{array}{cccc}
0 & 1 & 0 & 0 \\
\frac{1}{9} & 0 & \frac{4}{9}\mbox{e}^{ik} & \frac{4}{9} \\
\frac{4}{9} & 0 & \frac{4}{9}\mbox{e}^{ik} & \frac{1}{9} \\
\frac{4}{9}\mbox{e}^{-ik} & 0 & \frac{1}{9} & \frac{4}{9}\mbox{e}^{-ik}
\end{array}
\right]
\end{equation*}
where the columns and rows are arranged in the following order $\rm (b,\hat{b
},a,\hat{a})$. The matrix ${\mathsf Q}(s,k)$ is obtained by multiplication
with the diagonal matrix $\delta_{\rm bb^{\prime }} \mbox{e}^{-\frac{s}{v}l_{
\rm b^{\prime}}}$. The determinant in Eq. (\ref{Q.periodic}) can be
computed and gives
\begin{equation}
\det\left[{\mathsf I}-{\mathsf
Q}(s,k)\right]=1+\frac{5}{27}\mbox{e}^{-2\frac{s}{v}a}-\frac{1}{9}
\mbox{e}^{-2\frac{s}{v}g}+\frac{1}{9}\mbox{e}^{-2\frac{s}{v}(a+g)} -\frac{8}{9}
\mbox{e}^{-\frac{s}{v}a}\left( 1+\frac{\mbox{e}^{-2\frac{s}{v}g}}{3}\right)
\cos k
\label{detQ.example}
\end{equation}
where $a$ is the length of the bond $\rm a$ and
$g$ is the length of the bond $\rm b$. As we said the solutions of
$\det\left[{\mathsf I}-{\mathsf Q}(s,k)\right]=0$ gives the desired
functions $s_{j}(k)$. For
this example, we plot the first branches in Fig.
\ref{fig.ex.dif} where we observe that, indeed, only one branch includes
the point $s=0$ at
$k=0$. This unique branch can be identified with the dispersion relation of
the hydrodynamic mode of diffusion.

The diffusion coefficient is obtained from the second derivative of the
first branch at $k=0$. This can be analytically computed for this
particular example as follows. We consider $s<<{\rm Min}\lbrace v/a,
v/g\rbrace$ and $k<<1$ and
expand Eq. (\ref{detQ.example}). After some simple algebra, we get
\begin{equation*}
\det\left[{\mathsf I}-{\mathsf Q}(s,k)\right]=\frac{16}{27}\left[ \frac{s}{v}
(a+g) + k^{2}\right] +{\cal O}(s^2)+{\cal O}(sk^2)+{\cal O}(k^4)
\end{equation*}
from which we obtain that the diffusion coefficient defined by Eq. (\ref
{diff.coef}) is
\begin{equation}
D=\frac{v}{a+g}  \label{D ejem}
\end{equation}

In general, the diffusion coefficient has the units of $[L^{2}]/[T].$ This
is because the wave number has the units of $1/[L]$. Since we
have considered $k$ as a dimensionless number, the diffusion
coefficient has here the units of $1/[T]$. In this example the standard
units can be recovered by considering $k=\kappa a$ with a standard wave
number $\kappa$ where $a$ is the period at which the unit cell is repeated. In
these units, we would have obtained $\tilde{D}=v\frac{a^{2}}{a+g}$. But, in
general,
for a more complicated graph, there is no bond length that we can associate
with the
periodicity of the chain as in this example, which is the reason why we
consider the
dimensionless parameter $k$. In this sense we are considering the
space in units of the unit cell of the periodic chain.

\subsection{A Green-Kubo formula for the diffusion coefficient}

In the previous example, we have seen that the diffusion coefficient is
inversely
proportional to the total length of the unit cell. This is a general
property that follows from a general expression for the diffusion
coefficient that we shall now obtain. From now on we shall consider
one-dimensional
chains\footnote{  The theory developed here is trivially extended for
graphs that display
periodicity in higher dimensions by considering the appropriate
dimensionality for
the vector $\bf a$.}.  Accordingly, $k$ is a scalar wavenumber and no longer
a vector.

Consider the vector $\pmb{\chi}_{0,k}$ defined by Eq. (\ref
{eigen.periodic}) and the eigenvector $\tilde{\pmb{\chi}}_{0,k}$ of
the adjoint matrix ${\mathsf Q}\left[s_{0}(k),k\right]^{\dagger}$
defined by
\begin{equation}
{\mathsf Q}\left[s_{0}(k),k\right]^{\dagger}\cdot \tilde{\pmb{\chi}}_{0,k}
= \tilde{\pmb{\chi}}_{0,k}  \label{eigen.periodic'}
\end{equation}
or equivalently
\begin{equation}
\tilde{\pmb{\chi}}_{0,k}^{\dagger}=\tilde{\pmb{\chi}}_{0,k}^{\dagger}\cdot
{\mathsf Q}\left[s_{0}(k),k\right]
\label{eigen.periodic''}
\end{equation}
This eigenvector satisfies
\begin{equation}
\tilde{\chi}_{0,k}[{\mathrm{b}}]_{\ \overrightarrow{
k\rightarrow 0}}\ 1  \label{eigen.norm2}
\end{equation}
Such vectors are normalized as
\begin{equation}
(\tilde{\pmb{\chi}}_{0,k},\pmb{\chi}_{0,k})=\sum_{{\mathrm{b}}}
\tilde{\chi}_{0,k}[{\mathrm{b}}]^{\ast} \chi _{0,k}[
{\mathrm{b}}]=1  \label{norm.periodic}
\end{equation}
On the other hand, due to Eqs. (\ref{eigen.periodic}) and (\ref
{eigen.periodic'}), we have
\begin{equation}
\left(\tilde{\pmb{\chi}}_{0,k},{\mathsf Q}\left(s_{0},k\right)
\cdot \pmb{\chi}_{0,k}\right)=1.
\label{norm.periodic'}
\end{equation}
with $s_0=s_0(k)$.
Differentiating Eq. (\ref{norm.periodic}) and Eq. (\ref{norm.periodic'}) with
respect to $k$ we get respectively
\begin{equation*}
\left(\frac{d\tilde{\pmb{\chi}}_{0,k}}{dk},\pmb{\chi}_{0,k}
\right)+\left(\tilde{\pmb{\chi}}_{0,k},\frac{d\pmb{\chi}_{0,k}}{dk}\right)=0
\end{equation*}
and
\begin{equation*}
\left(\frac{d\tilde{\pmb{\chi}}_{0,k}}{dk},\pmb{\chi}_{0,k
}\right)+\left(\tilde{\pmb{\chi}}_{0,k},\left[ \frac{d}{dk}{\mathsf Q}(
s_{0},k)\right] \cdot\pmb{\chi}_{0,k}\right)+\left(\tilde{\pmb{\chi}}
_{0,k},\frac{d\pmb{\chi}_{0,k}}{dk}\right)=0
\end{equation*}
where we have used Eqs. (\ref{eigen.periodic}) and (\ref{eigen.periodic'}).
These last two equations imply
\begin{equation}
\left(\tilde{\pmb{\chi}}_{0,k},\left[ \frac{d}{dk}{\mathsf Q}(
s_{0},k) \right] \cdot \pmb{\chi}_{0,k} \right) = \sum_{{\mathrm{b,b}}
^{\prime }} \tilde{\chi}_{0,k}[{\mathrm{b}}]^{\ast } \left[ \frac{d
}{dk} Q_{{\mathrm{bb}}^{\prime }} (s_{0},k)
\right] \chi_{0,k}[{\mathrm{b}}^{\prime }]=0
\label{eq.Qderiv}
\end{equation}
Now we compute the derivative of $\mathsf Q$. From Eq. (\ref{defQ}) and
Eq. (\ref{Trans.periodic}), we have
\begin{equation*}
\frac{d}{dk}Q_{{\mathrm{bb}}^{\prime }}( s_{0}, k ) =\left[
\frac{d}{dk}{\mathrm{P}}_{{\mathrm{bb}}^{\prime }}(k)\right] \exp
\left(-s_0\frac{l_{{\mathrm{b}}^{\prime }}  }{v}\right) -
Q_{{\mathrm{bb}}^{\prime}}
(s_{0},k) \frac{  l_{{\mathrm{b}}^{\prime }}}{v}\frac{ds_{0}}{dk}.
\end{equation*}
Inserting this result into Eq. (\ref{eq.Qderiv}) and taking the limit $k
\rightarrow 0$, we obtain
\begin{equation*}
\left. \frac{ds_{0}}{dk}\right| _{k=0}=\frac{\frac{1}{2B}
\sum_{{\mathrm{b,b}}^{\prime }}\left.
\frac{d{\mathrm{P}}_{{\mathrm{bb}}^{\prime }}
}{dk}\right| _{k=0}}{\frac{1}{2B}\sum_{{\mathrm{b,b}}
^{\prime }}\left. {\mathrm{P}}_{{\mathrm{bb}}^{\prime }}\right| _{k=0}
\frac{l_{{\mathrm{b}}^{\prime }}}{v}},
\end{equation*}
where we have used the fact that the limit $k\rightarrow 0$ implies $s_{0}(
k)\rightarrow 0$, and that $\chi _{0,k}[{\mathrm{b}}^{\prime }]\rightarrow
\frac{1
}{2B},\tilde{\chi }_{0,k}[{\mathrm{b]}\rightarrow 1}$ because of Eqs.
(\ref{eigen.norm}) and
(\ref {eigen.norm2}).  Since
$\sum_{{\mathrm{b}}}\left. {\mathrm{P}}_{{\mathrm{bb}}^{\prime
}}\right|_{k=0} =
1$, $\forall \,{\mathrm{b}}^{\prime }$, this reduces to:
\begin{equation}
\left. \frac{ds_{0}}{dk}\right| _{k=0}=\frac{v}{L
_{{\mathrm{uc}}}} \sum_{{\mathrm{b,b}}^{\prime }}\left. \frac{d{\mathrm{P}}_{
{\mathrm{bb}}^{\prime }}}{dk}\right| _{k=0}=0
\label{eq.der.s}
\end{equation}
where $L_{{\mathrm{uc}}}=\sum_{{\mathrm{b}}}l_{{\mathrm{b}}}$ is the
total length of the unit cell.  The last equality $(=0)$ follows from the fact
that the unit cell is connected with the neighboring cells in a symmetric way.
For instance, in a one-dimensional chain, the ``fluxes'' from the left-hand
side equal
those from the right-hand side and the derivative with respect to $k$ drops a
sign that makes the sum vanishing. The reader can verify this property in
the previous example of the comb graph.

To obtain the diffusion coefficient we need the second derivative of $s_{0}(
k)$.  Therefore, we differentiate Eq. (\ref{eq.Qderiv}) with respect to
$k$ and we evaluate at $k=0$. After some algebra and using
Eq. (\ref{eq.der.s}), we get
\begin{equation}
\left. \frac{d^{2}s_{0}}{dk^{2}}\right| _{k=0}=\frac{v}{
L_{\mathrm{uc}}}\left[ \sum_{\mathrm{b,b}^{\prime }}\left. \frac{
d^{2}\mathrm{P}_{\mathrm{bb}^{\prime }}}{dk^{2}}\right| _{
k=0}+\sum_{\mathrm{b,b}^{\prime }}\left. \frac{d\mathrm{P}_{\mathrm{bb}
^{\prime }}}{dk}\left( 2B\frac{d\chi _{0,k}[\mathrm{b}
^{\prime }\mathrm{]}}{dk}+\frac{d\tilde{\chi }_{0,k
}[\mathrm{b}]^{\ast }}{dk}\right) \right| _{k=0}\right]
\label{diff}
\end{equation}

The explicit form for the diffusion coefficient is obtained from Eq. (\ref
{diff}) if we compute the first derivatives of the eigenstates. We can write
the equations that these quantities satisfy.  In fact taking the derivative
with respect to $k$ of Eqs. (\ref{eigen.periodic}) and (\ref
{eigen.periodic''}) we have
\begin{equation*}
\frac{d\chi_{0,k}[{\mathrm{b}}]}{dk}=\sum_{{\mathrm{b}}
^{\prime }}\frac{dQ_{{\mathrm{bb}}^{\prime }}}{dk}\chi_{0,
k}[{\mathrm{b}}]+\sum_{{\mathrm{b}}^{\prime }} Q_{{\mathrm{bb}}
^{\prime }}\frac{d\chi _{0,k}[{\mathrm{b}}^{\prime }]}{dk}
\end{equation*}
and
\begin{equation*}
\frac{d\tilde{\chi }_{0,k}[{\mathrm{b}}^{\prime }]^{\ast }
}{dk}=\sum_{{\mathrm{b}}}\tilde{\chi }_{0,k}[
{\mathrm{b}}]^{\ast }\frac{dQ_{\mathrm{bb}^{\prime }}}{dk}+\sum_{
\mathrm{b}}\frac{d\tilde{\chi }_{0,k}[{\mathrm{b}}]^{\ast }}{d
k}Q_{\mathrm{bb}^{\prime }}
\end{equation*}
whose solutions are
\begin{equation*}
\frac{d\chi _{0,k}[{\mathrm{b}}]}{dk}=\sum_{\mathrm{
b^{\prime }}\mathrm{b^{\prime \prime }}}\left({\mathsf I}-{\mathsf
Q}\right)_{\mathrm{bb}^{\prime
}}^{-1}\frac{dQ_{\mathrm{b^{\prime }b}^{\prime \prime }}}{dk}
\chi_{0,k}[\mathrm{b}^{\prime \prime }]
\end{equation*}
\begin{equation*}
\frac{d\tilde{\chi }_{0,k}[{\mathrm{b}}]^{\ast }}{d
k}=\sum_{\mathrm{b^{\prime }}\mathrm{b^{\prime \prime }}}\tilde{
\chi }_{0,k}[{\mathrm{b}}^{\prime }]^{\ast }\frac{dQ_{
\mathrm{b^{\prime }b}^{\prime \prime }}}{dk}\left({\mathsf I}-{\mathsf
Q}\right)_{\mathrm{
b^{\prime \prime }b}}^{-1}
\end{equation*}

In the limit $k\rightarrow 0$, these solutions can be written as
\begin{equation*}
\left. \frac{d\chi_{0,k}[{\rm b}]}{dk} \right|_{k=0}
=\frac{1}{2B} \sum_{ {\rm b}^{\prime},{\rm b}^{\prime \prime }}
\sum_{n=0}^{\infty}
\left[
\left({\mathsf P}^{n}\right)
_{\mathrm{b,b^{\prime }}}
\frac{d{\mathrm P}_{{\rm b}^{\prime },{\rm b}^{\prime \prime}}}{dk}
\right]_{k=0}
\end{equation*}
and similarly
\begin{equation*}
\left. \frac{d\tilde{\chi }_{0,k}[{\rm b}]^{\ast}}{dk} \right| _{k=0}
=\sum_{{\rm b}^{\prime },{\rm b}^{\prime \prime }}
\sum_{n=0}^{\infty }
\left[ \frac{ d{\mathrm{P}}
_{{\rm b}^{\prime},{\rm b}^{\prime \prime }}}{dk}
\left( {\mathsf P}^{n}\right)_{{\rm b}^{\prime \prime },{\rm b}}\right]_{k=0}
\end{equation*}

Thus, the second term of Eq. (\ref{diff}) becomes
\begin{multline*}
\sum_{\mathrm{b,b}^{\prime }}\left. \frac{d\mathrm{P}_{\mathrm{bb}^{\prime }}
}{dk}\left( 2B\frac{d\chi _{0,k}[\mathrm{b}^{\prime }
]}{dk}+\frac{d\tilde{\chi }_{0,\mathrm{k}}[
\mathrm{b}]^{\ast }}{dk}\right) \right| _{k=0} \\ =
2\sum_{\mathrm{b,b^{\prime },b^{\prime \prime },b^{\prime \prime \prime }}
}\sum_{n=0}^{\infty }\left[ \frac{d{\mathrm{P}}_{\mathrm{bb^{\prime }}}
}{dk}\left( {\mathsf P}^{n}\right) _{\mathrm{b^{\prime },b^{\prime
\prime }}}\frac{d\mathrm{P}_{\mathrm{b^{\prime \prime }b^{\prime \prime
\prime }}}}{dk}\right]_{k=0}
\end{multline*}
To evaluate and interpret this result we have to transform these
expressions. First, we have to consider the derivatives of ${\mathsf P}(k)$.
This matrix is defined in Eq. (\ref{Trans.periodic}). Since only the nearest
neighbors are connected, the lattice vector of the jumps can take only the
values $a({\rm
b},{\rm b}^{\prime})=0,\pm 1$ whether the particle crosses the boundary of
the unit cell to the
right-hand cell (+1), or the left-hand one (-1), or it stays in the same
cell (0)
during the transition ${\rm b}'\to{\rm b}$. Therefore
\begin{equation*}
{\mathrm P}_{{\rm bb}^{\prime }}(k) =P_{T_{a({\rm b},{\rm b}')}({\rm b}),
{\mathrm b^{\prime }}}\ \mbox{e}^{-ik a({\rm b},{\rm b}')}
\end{equation*}
The derivatives of this matrix are thus
\begin{equation*}
\frac{d{\mathrm{P}}_{\mathrm{bb^{\prime }}}}{dk}=-i\; a({\rm b},{\rm
b}^{\prime})
\ {\mathrm P}_{{\rm bb}^{\prime }}
\end{equation*}
and
\begin{equation*}
\frac{d^{2}{\mathrm{P}}_{\mathrm{bb^{\prime }}}}{dk^{2}}=-a({\rm b},{\rm
b}^{\prime})^2
\ {\mathrm P}_{{\rm bb}^{\prime }}.
\end{equation*}
Accordingly, the diffusion coefficient is given by
\begin{multline}
D=-\frac{1}{2}
\left. \frac{d^{2}s_{0}}{dk^{2}}\right| _{k=0}=\frac{v}{2
{L}_{\mathrm{uc}}}\sum_{\mathrm{b,b^{\prime }}}\biggl\lbrack a({\rm b},{\rm
b}^{\prime})^{2}
{\mathrm P}_{\mathrm{bb^{\prime }}} \\
 +2\sum_{\mathrm{b^{\prime \prime }b^{\prime\prime
\prime }}}\sum_{n=0}^{\infty } a({\rm b},{\rm b}^{\prime}) {\mathrm
P}_{\mathrm{bb^{\prime
}}} \left({\mathsf P}^n\right)_{\mathrm{b^{\prime },b^{\prime \prime }} }
a({\mathrm{b^{\prime
\prime}}},{\mathrm{b^{\prime \prime\prime}}}) {\mathrm
P}_{\mathrm{b^{\prime \prime }b^{\prime \prime
\prime }}} \biggr\rbrack_{k=0}
\label{diff.1}
\end{multline}
In order to interpret this formula, we have to remember some definitions.
If an observable is
defined over $M$ successive bonds, its mean value over the equilibrium
invariant measure of the
random process is given by
\begin{equation}
\left\langle A\right\rangle =\lim_{N\rightarrow \infty }\sum_{b_{-N}\cdots
b_{N}} A(b_{M-1}\cdots b_{1}b_{0}) \; \mu(b_{N}\cdots b_{-N}) .
\label{def.mean}
\end{equation}
If the observable $A$ depends only on two consecutive bonds as it is the
case for the jump vector
$a({\rm b},{\rm b}')$, its mean value takes the form
\begin{equation*}
\left\langle A\right\rangle =\sum_{b_0b_1} A(b_1b_0) \; \mu(b_1b_0) =
 \sum_{b_0b_1}A(b_1b_0)\; P_{b_1b_0} \; \chi_0[b_0] =
\frac{1}{2B} \sum_{b_0b_1}A(b_1b_0)\; P_{b_1b_0}
\end{equation*}
because of Eqs. (\ref{inv.meas}) and (\ref{eigen.norm}).
According to the general definition (\ref{def.mean}), the time-discrete
autocorrelation function of
a two-bond observable is given by
\begin{equation*}
\left\langle A_{m}A_{0}\right\rangle =\sum_{b_{0}\cdots
b_{m+1}}A(b_{m+1}b_m)A(b_1b_0)\; \mu(b_{m+1}b_m\cdots b_{1}b_0)
\end{equation*}
Because of Eqs. (\ref{inv.meas}) and (\ref{eigen.norm}) again, we get
\begin{equation}
\left\langle A_{m}A_{0}\right\rangle =\frac{1}{2B}\sum_{b_0 b_1 b_m
b_{m+1}} A(b_{m+1}b_m)
P_{b_{m+1}b_m} \left({\mathsf P}^{m-1}\right)_{b_mb_1} A(b_1b_0)\; P_{b_1b_0}
\label{correl}
\end{equation}
The terms of Eq. (\ref{diff.1}) are precisely of the form of Eq.
(\ref{correl}) with
$m=0$ for the first term and $m=n+1$ for the following ones, and with the
observable $A=a$.
Since the process is stationary we have that $\left\langle
a_{n}a_{0}\right\rangle =\left\langle
a_{0}a_{-n}\right\rangle =\left\langle a_{-n}a_{0}\right\rangle $ where the
last equality follows from the commutativity of the quantities $a_0$ and
$a_{-n}$. Therefore, the
term with the sum over $n$ in Eq. (\ref{diff.1}) is equal to
$2B\sum_{n=-\infty ,n\neq 0}^{+\infty
}\left\langle a_{0}a_{n}\right\rangle$. It is now clear that Eq. (\ref
{diff.1}) for the diffusion coefficient is

\begin{equation}
D=\frac{Bv}{L_{\mathrm{uc}}} \sum_{n=-\infty }^{+\infty
}\left\langle a_{0}a_{n}\right\rangle =
\frac{v}{2\left\langle l\right\rangle} \sum_{n=-\infty }^{+\infty
}\left\langle a_{0}a_{n}\right\rangle
\label{diff.2}
\end{equation}
where $\left\langle l\right\rangle=L_{\rm uc}/(2B)$ is the mean bond-length
of the unit cell and
$a_{n}=0,\pm 1$ is the jump from one cell to another undergone by the
particle in motion on the
infinite graph.

Eq. (\ref{diff.2}) is nothing else than the Green-Kubo formula for the
diffusion coefficient. If we define
\begin{equation*}
v_{x}=\frac{\Delta x}{\Delta t}= \frac{v}{\left\langle l\right\rangle} \;
a({\rm b},{\rm b}')
\end{equation*}
as the velocity along the $x$-axis that contributes to the transport,
where $\Delta t=\left\langle l\right\rangle/v$, we can write Eq. (\ref{diff.2})
in the more familiar Green-Kubo form
\begin{equation*}
D=\frac{1}{2}\int_{-\infty }^{+\infty }\left\langle
v_{x}(0)v_{x}(t)\right\rangle dt.
\end{equation*}

In the time-discrete form (\ref{diff.2}), we obtain the result that the
diffusion coefficient is proportional to the constant velocity $v$ and
inversely proportional to the
mean bond-length of a unit cell. The diffusion coefficient is
also proportional to the sum of the time-discrete autocorrelation of the
jump $a$ from cell to cell.

\section{Escape and diffusion on large open graphs}

\label{sec.esc.diff}

\begin{figure}[th]
\centering
\includegraphics[width=12cm]{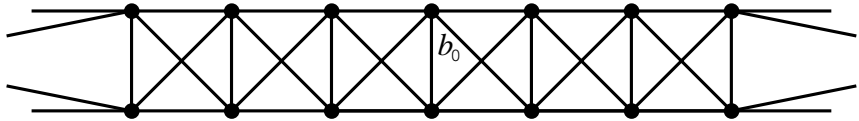}
\caption{Open graph forming a chain composed of $N=6$ identical
unit cells. The five lengths that compose the unit cell take
different values. (In this regard, no specific length
can be associated with the periodicity of the chain.)}
\label{fig.chain}
\end{figure}

In this section, we shall study the Pollicott-Ruelle resonances of open
graphs characterized by a unit cell which is repeated a finite
number of times. The particular example that we consider is depicted in
Fig. \ref{fig.chain}. We shall focus on the leading resonance that
determines the escape rate from the system.  We shall show that, for large
enough chains
 (i.e., made of several unit cells), the classical lifetime corresponds to the
time spent by a particle that undergoes a diffusive process in the chain
before it
escapes.

For the graph of Fig. \ref{fig.chain}, the transition probabilities from
bond to bond $P_{bb^{\prime }}$ are given by
\begin{equation*}
P_{bb^{\prime }}=\left\{
\begin{array}{l}
\frac{9}{25}\text{ if the particle is reflected, i.e., }b=b^{\prime }\; ; \\
\frac{4}{25}\text{ for bonds }b\neq b^{\prime }\text{ which are
connected}\; ; \\
0\ \text{  otherwise}\; .
\end{array}
\right.
\end{equation*}
We have computed the spectrum of Pollicott-Ruelle resonances for different
values
of the number $N$ of unit cells.  The leading resonance controls the asymptotic
decay.  Since the leading resonance is isolated and at a finite distance
from the real axis the decay is exponential as we explained, that is
\begin{equation*}
\rho (t)\sim \exp\left(-\gamma_{\rm cl}t\right)
\end{equation*}
where $\gamma_{\rm cl}$ is the leading resonance, i.e.,
the escape rate. This is the
generic behavior of the density in a classically chaotic open system and we
refer to this as the classical decay.

\begin{figure}
\centering
\includegraphics[width=9cm]{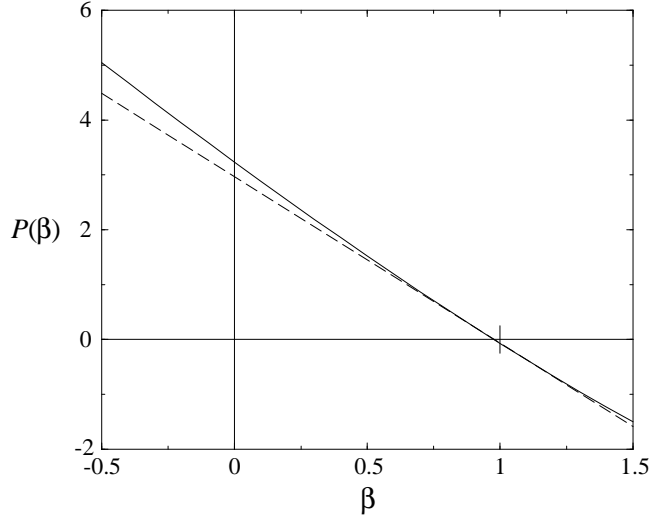}
\caption{Topological pressure for the chain of Fig. \ref{fig.chain} but
with $N=7$.
From this function, we get that $h_{\rm KS} \approx 2.9>0$
and $h_{\rm top}=3.2330$.}
\label{fig.presN7}
\end{figure}

In Fig. \ref{fig.presN7}, we depict
the topological pressure for the chain of Fig. \ref{fig.chain} with $N=7$
unit cells.

When the chaotic dynamics is at the origin of a diffusion process the
escape rate
is inversely proportional to the square of the size of the system, more
precisely the following relation is expected to hold
\begin{equation}
\gamma_{\rm cl}(N) \simeq D\; \frac{\pi ^{2}}{N^{2}}  \label{gamma.diff}
\end{equation}
with $D$ the diffusion coefficient. \footnote{
This relation is obtained by solving the diffusion equation
(\ref{diffusion.eq})
in a system of size $N$ with absorbing boundary conditions at the
borders, i.e, $\rho (0)=0$ and $\rho (N)=0$.  The mode with the slowest
decay is then given by $\sin (kx)$ with $k=\frac{\pi }{N}$ and,
from the dispersion relation (\ref{diff.disp.rel}), we get Eq. (
\ref{gamma.diff}). For large systems, when $N\rightarrow \infty$,
or equivalently $k\rightarrow 0$,
$D(N)\equiv\gamma_{\rm cl}(N)(N/\pi)^2$ must approach the diffusion
coefficient.}
As explained in Sec. \ref{sec.chaotic}, the escape rate is related to the
mean Lyapunov exponent and
the KS entropy of the open chain of size $N$ according to
\begin{equation}
\gamma_{\rm cl}(N) = \lambda(N) - h_{\rm KS}(N)
\end{equation}
As a consequence of Eq. (\ref{gamma.diff}), we find also for open graphs a
known relationship between
the diffusion coefficient and the chaotic properties \cite{GaspNicolis}
\begin{equation}
D = \lim_{N\to\infty} \frac{N^2}{\pi^2} \; \left[\lambda(N) - h_{\rm
KS}(N)\right]
\end{equation}

Here again $D$ is in units of $[1]/[T]$ because we did not associate a length
with the period of the chain and thus the space is in units of unit cell.

\begin{figure}[ht]
\centering
\includegraphics[width=6cm]{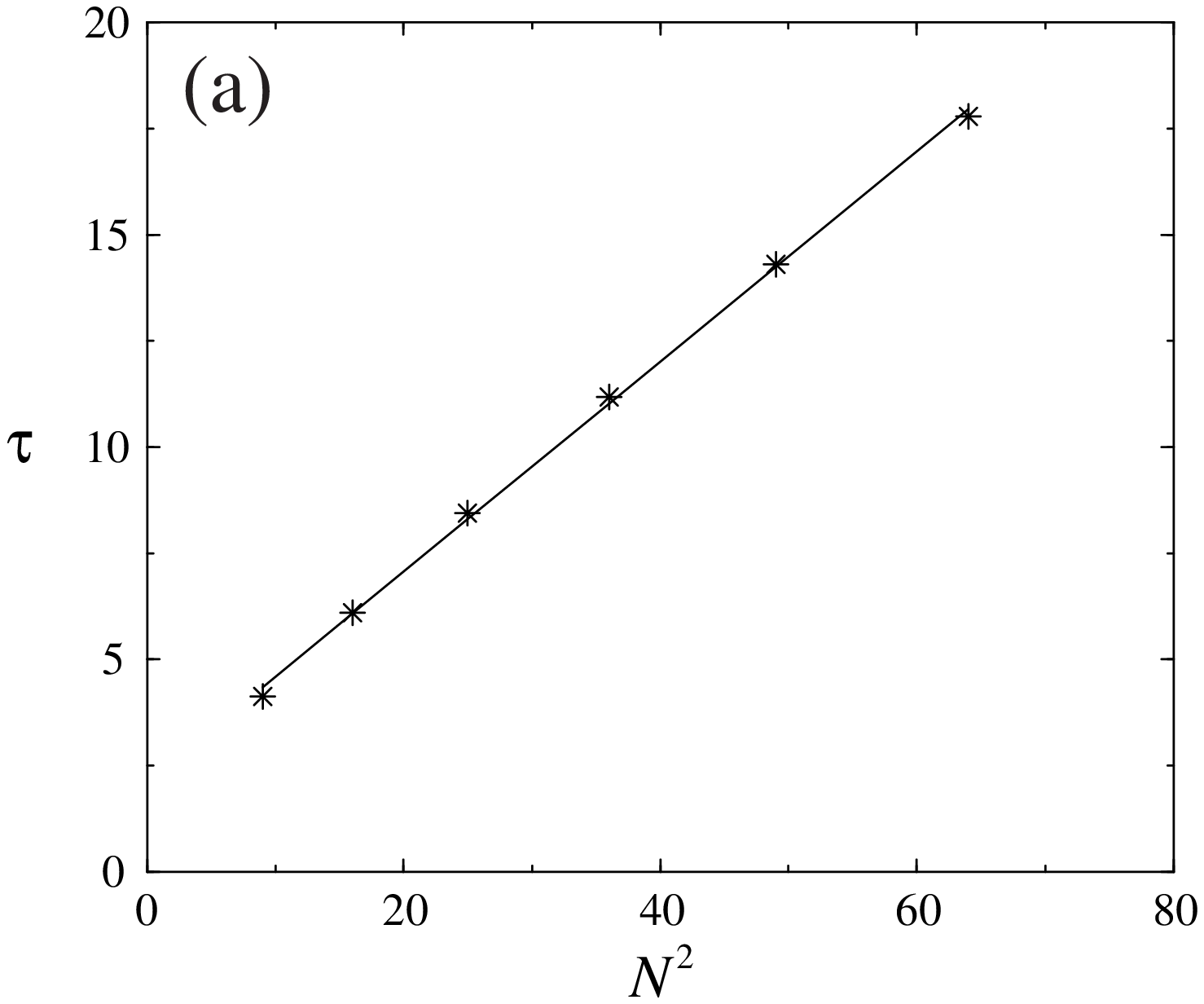}
\includegraphics[width=6.3cm]{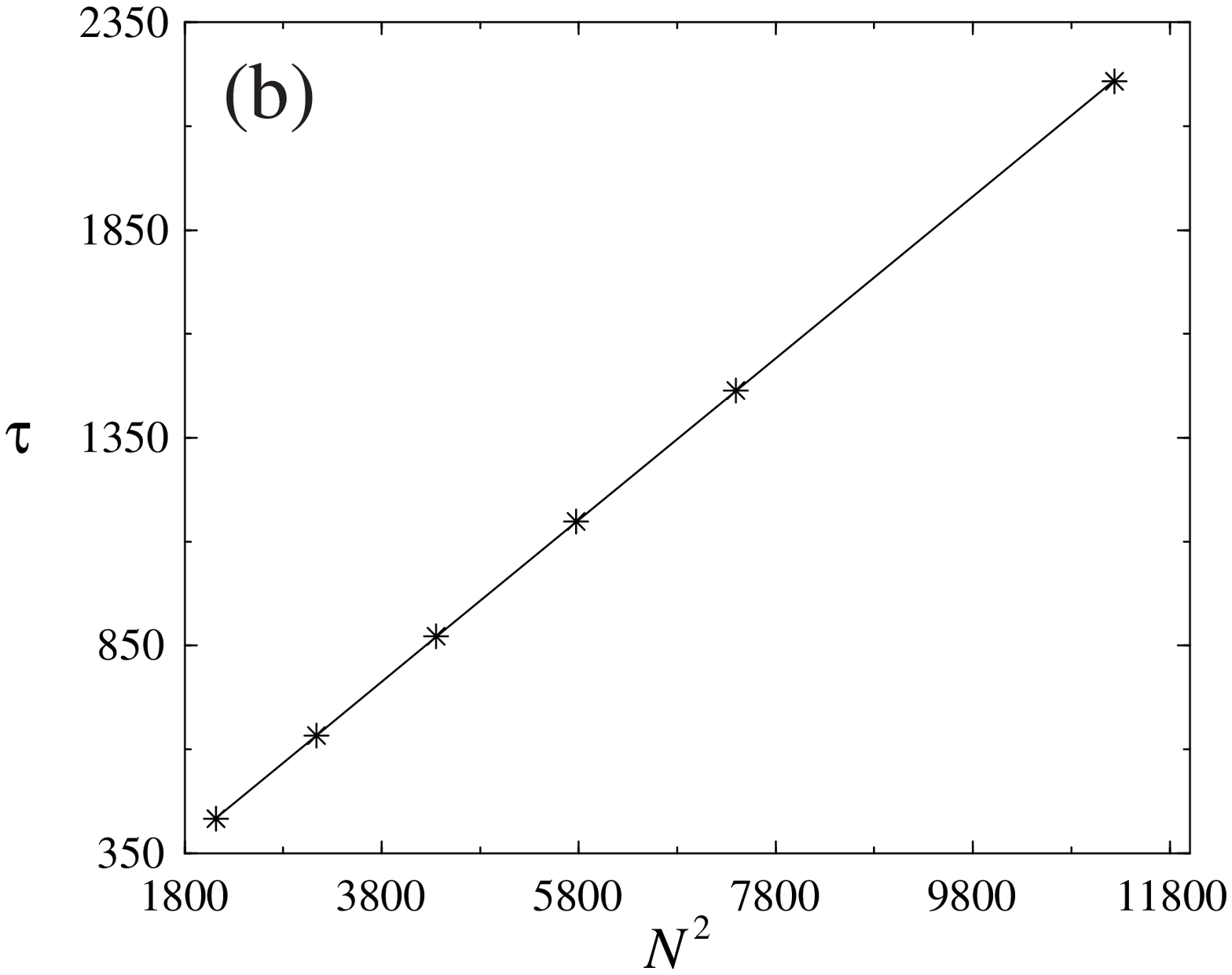}
\caption{Lifetime of the chain of Fig. \ref{fig.chain} as a function of the
square of its size $N$:
 (a) For $N=3,4,5,6,7,8$, the slope of the linear regression is $0.247403$
and Eq. (\ref{gamma.diff}) gives an approximate diffusion coefficient $D
\simeq 0.4095$.
(b) For $N=46,56,66,76,86,106$, the slope of the linear regression
is $0.1946247$ and Eq. (\ref{gamma.diff}) gives a better approximation $D
\simeq 0.5205$ for the
diffusion coefficient (\ref{The.diff.coeff}).}
\label{fig.diff}
\end{figure}

We have computed the escape rate for chains of different sizes. If the
dynamics indeed corresponds to a diffusion process, then Eq. (\ref{gamma.diff})
should be verified. In figure \ref{fig.diff}, we plot the classical
lifetimes $\tau_{\rm cl}$ as a function of $N^2$. Since $\tau_{\rm cl}
=1/\gamma_{\rm cl}$, we
observe the dependence of $\gamma_{\rm cl}$ on $N$ expected from Eq.
(\ref{gamma.diff}).  We may conclude from this result that the classical
dynamics in the open
chain is the one of a diffusion process.

\begin{figure}[ht]
\centering
\includegraphics[width=9cm]{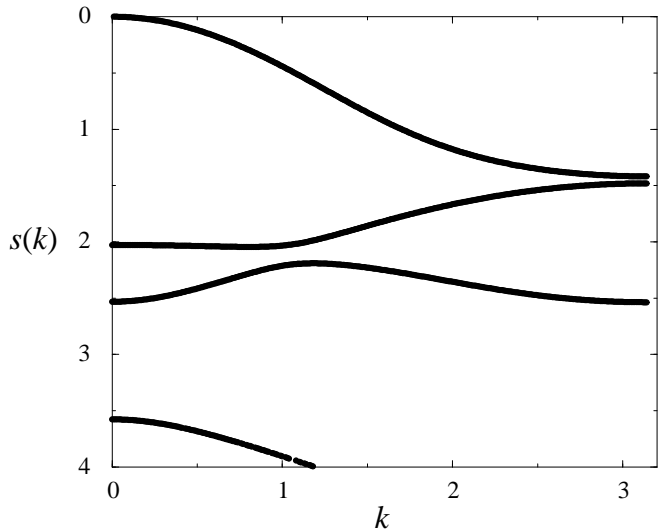}
\caption{The first branches $s(k)$ of Pollicott-Ruelle resonances of the
infinite graph
corresponding to the open graph of Fig. \ref{fig.chain}. Here again we
observe that there is only
one branch that is identified with the hydrodynamic mode of diffusion,
$s_0(k)$, because it vanishes for $k=0$.}
\label{fig.branches}
\end{figure}

The diffusion coefficient of the infinite graph can be obtained as we explained
in Sec. \ref{sec.diff}. We depict in Fig. \ref{fig.branches} the leading
and other Pollicott-Ruelle
resonances of the infinite graph obtained by numerical calculation as a
function of the
dimensionless wave number $k$. The diffusion coefficient is given by the
second derivative of the
leading resonance $s_{0}(k)$ evaluated at $k=0$. In this way, we obtain the
numerical result
\begin{equation}
D =0.5318.
\label{The.diff.coeff}
\end{equation}
Accordingly, the diffusion coefficient of the infinite
chain gives a reasonable estimate for the proportionality coefficient
between $\gamma_{\rm cl}$ and $1/N^{2}$ for the small chains of Fig.
\ref{fig.diff}a
and is a very good estimate for the large chains of Fig. \ref{fig.diff}b.

\begin{figure}[ht]
\centering
\includegraphics[width=10cm]{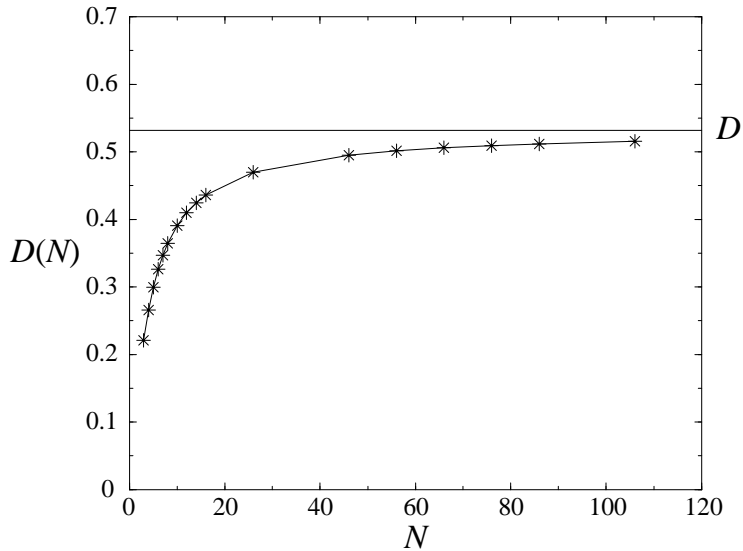}
\caption{The effective diffusion coefficient $D(N)=\gamma_{\rm
cl}(N)(N/\pi)^2$ as a function of $N$
for the open graphs of Fig. \ref{fig.chain}.}
\label{fig.dif.conv}
\end{figure}

In Fig. \ref{fig.dif.conv}, we show how the effective diffusion coefficient
$D(N)=\gamma_{\rm
cl}(N) (N/\pi)^2$ converges to the diffusion coefficient of the infinite
chain $D$, as
the chain becomes longer and longer ($N\to\infty$).

\section{Diffusion in disordered graphs}
\label{sec.disorder}

In a recent work, Schanz and Smilansky \cite{Schanz} have considered the
problem of Anderson
localization in a one-dimensional graph composed of successive bonds of
random lengths with random
transmission and reflection coefficients at the vertices.  The classical
dynamics corresponding to
this quantum model defines a kind of Lorentz lattice gases as studied in
Refs. \cite{Ernst,Appert}.
Indeed, these references describe Lorentz lattice gases consisting of a
moving particle traveling
with allowed velocities $\pm v$ on a one-dimensional lattice of scatterers.
If the particle arrives
at a scatterer it will be transmitted or reflected with probabilities $p$
and $q=1-p$
respectively.  If the scatterers are randomly distributed the model
describes the classical dynamics
of the model by Schanz and Smilansky with identical transmission and
reflection coefficients at all
the scatterers.

The classical dynamics of this model can be analyzed with the methods
developed in the
present paper, which provides the relationship with the cited works on the
Lorentz lattice gases.
Using the methods of Sec. \ref{sec.relax}, we can write down the infinite
matrix ${\mathsf
Q}(s)$.  The eigenstates of this matrix corresponding to the unit
eigenvalue can be obtained by
iteration along the chain according to $\chi[b]=\exp(sl_b/2v)u_b$ and
$\chi[\hat{b}]=\exp(sl_b/2v)u_{\hat b}$ with
\begin{equation}
\left(
\begin{array}{c} u_{\hat{b}} \\ u_{b}
\end{array}
\right)
={\mathsf M}_{b} \cdot
\left(\begin{array}{c} u_{\widehat{b-1}} \\ u_{b-1}
\end{array}\right)
\label{iter}
\end{equation}
where $b\in{\mathbb Z}$ and the matrix is defined by
\begin{equation}
{\mathsf M}_{b} = {\mathsf B}_{b}^{1/2} \cdot {\mathsf V}_{b-1} \cdot
{\mathsf B}_{b-1}^{1/2}
\end{equation}
with
\begin{equation}
{\mathsf V}_{b-1} =
\left(
\begin{array}{cc} \frac{1}{p_{b}} & -\frac{q_{b}}{p_{b}} \\
\frac{q_{b}}{p_{b}} &
1-\frac{q_{b}}{p_{b}}
\end{array}
\right)
\end{equation}
and
\begin{equation}
{\mathsf B}_b =
\left(
\begin{array}{cc} \mbox{e}^{+\frac{sl_b}{v}} & 0 \\
0 & \mbox{e}^{-\frac{sl_b}{v}}
\end{array}
\right)
\end{equation}
We notice that $\det{\mathsf M}_b=1$.  If the chain was periodic $l_b=l$,
we would obtain the
diffusion coefficient $\tilde{D}= v l p/(2q)$ by assuming that
$u_{b+1}=\exp(i\kappa l)u_b$
in Eq. (\ref{iter}). In the dilute gas limit, the mean-field diffusion
coefficient for the random
graph is then given by replacing the bond length $l$ by the mean bond
length $\langle l\rangle$,
leading to:
\begin{equation}
\tilde{D}_{\rm mf}= v \; \langle l\rangle  \; \frac{p}{2q}
\label{diff.disorder}
\end{equation}

For a disordered chain with $N$ scatterers,
the Pollicott-Ruelle resonances can be obtained by finding the resonances
$s$ for which the
following equation is satisfied:
\begin{equation}
\left(
\begin{array}{c} u_{\widehat{N+1}} \\ u_{N+1}
\end{array}
\right)
=\prod_{b=1}^{N}{\mathsf M}_{b}(s) \cdot
\left(\begin{array}{c} u_{\hat{1}} \\ u_{1}
\end{array}\right)
\end{equation}
If the chain closes on itself, we must impose the periodic boundary
conditions $u_{\widehat
{N+1}}=u_{\hat 1}$ and $u_{N+1}=u_1$.  If the chain is open and extended by two
semi-infinite leads, we must consider the absorbing boundary conditions
$u_{\widehat
{N+1}}=0$ and $u_{1}=0$.

In Ref. \cite{Ernst}, Ernst {\em et al.} have characterized the chaotic
properties in such open
graphs thanks to the escape-rate formalism by computing the topological
pressure function of Sec.
\ref{sec.chaotic}.  In Ref. \cite{Appert}, Appert {\em et al.} showed that
the spatial disorder is
at the origin of a dynamical phase transition associated with a singularity
in the pressure function
of the infinite disordered chain.

\section{Conclusions}
\label{conclu}

In this paper, we have introduced and studied the random classical dynamics
of a particle moving in
a graph. We shall show elsewhere \cite{Felipe4} that the dynamics here
studied is the
classical limit of the quantum dynamics introduced in Refs.
\cite{Smilansky0,Smilansky1}.

We have shown that the relaxation rates of the time-continuous classical
dynamics can be obtained by
a simple secular equation which includes the lengths of the bonds and the
velocity of the particle.
This secular equation has been directly related to the eigenvalue
problem of the time-continuous Frobenius-Perron operator.  The secular
equation can be written as a
classical zeta function defined as a product over the periodic orbits of
the graphs.   In this way,
we have been able to define the relaxation rates, as well as chaotic
properties such as the Lyapunov
exponents and the entropies as quantities per unit of the continuous time.
The chaotic
properties are derived from a pressure function defined for each graph.

For infinite periodic graphs, we have shown how to construct the
hydrodynamic modes of diffusion
and to compute a diffusion coefficient.  Here also, the relaxation rates of
the hydrodynamic modes
are given by the zeros of a classical zeta function.  Moreover, a
Green-Kubo formula for the
diffusion coefficient has been deduced from the eigenvalue problem for the
Frobenius-Perron operator
of the classical dynamics on the graph.

When the chain is open by considering a finite segment connected with
scattering leads, the particle escapes after a diffusion process.  In this
case, we have computed the
lifetime of the metastable states.  This classical lifetime is given by the
inverse of the
escape rate which is related to the diffusion coefficient.  Accordingly, a
known relationship between
the diffusion coefficient and the chaotic properties \cite{GaspNicolis} is
extended to the random
classical dynamics on graphs.  The case of infinite disordered graphs has
also be considered.

The interest of these results lies notably in the fact that the classical
quantities here computed
can be compared to equivalent quantities defined for the corresponding
quantum problem, as shown
elsewhere \cite{Felipe4}.

%
\section*{Acknowledgments}
The authors thank Professor G. Nicolis for support and
encouragement in this research.  FB is financially supported by the
``Communaut\'e fran\c caise de Belgique" and PG by the National Fund
for Scientific Research (F.~N.~R.~S. Belgium).  This research is supported,
in part, by the Interuniversity Attraction Pole program of the
Belgian Federal Office of Scientific, Technical and Cultural Affairs,
and by the F.~N.~R.~S. .


\end{document}